\documentclass[12pt]{article}
\usepackage{epsf,epsfig,psfig,float,amssymb,latexsym}
\usepackage{graphics,psfrag}
\newcommand{\cR}{{\cal R}}

\newcommand{\Opd}{{\mathcal{O}}(p^2)}
\newcommand{\Opt}{{\mathcal{O}}(p^3)}
\newcommand{\Opc}{{\mathcal{O}}(p^4)}

\newcommand{\be}{\begin{equation}}
\newcommand{\ee}{\end{equation}}
\newcommand{\ba}{\begin{eqnarray}}
\newcommand{\ea}{\end{eqnarray}}

\newcommand{\nn}{\nonumber}

\newcommand{\vs}{\vspace{-0.20cm}}

\begin{document}

\thispagestyle{empty}

\vspace{2cm}

\begin{center}
{\Large{\bf Nucleon-Nucleon interactions from effective field theory}}
\end{center}
\vspace{.5cm}

\begin{center}
{\Large Jos\'e A. Oller\footnote{email: oller@um.es}}
\end{center}

\begin{center}
{\it {\it Departamento de F\'{\i}sica. Universidad de Murcia.\\ E-30071,
Murcia. Spain.}}
\end{center}
\vspace{1cm}

\begin{abstract}
\noindent
 We have established a new convergent scheme to treat
 analytically nucleon-nucleon interactions from a chiral effective field theory.
 The Kaplan-Savage-Wise (KSW) amplitudes are resummed to fulfill the unitarity 
 or right hand cut to all orders below pion production threshold. This is achieved 
 by matching  order by order in the KSW power counting the general expression
 of a partial  wave  with resummed unitarity cut, with the 
 inverses of the KSW amplitudes. As a result, a new convergent and systematic KSW 
 expansion is derived for an on-shell interacting kernel $\cR$ in terms of which the 
 partial waves are computed. The  
 agreement with data for the S-waves is fairly good up to laboratory 
 energies around 350 MeV and clearly improves and reestablishes 
 the phenomenological success of the KSW amplitudes when treated within 
 this scheme. 
\end{abstract}

\vspace{2cm}


\newpage

\section{Introduction}
\label{sec:intro}
\def\theequation{\arabic{section}.\arabic{equation}}
\setcounter{equation}{0}

Effective field theories are the standard method to deal with strong
interactions in the non-perturbative regime. As paradigm we have 
SU(2) Chiral Perturbation Theory (CHPT) for pion physics \cite{wein,gl,pich},  
where a convergent power counting is established in terms of derivatives, 
insertions of quark mass matrix and external sources. This has also been 
applied to  pion-nucleon interactions with baryon number, B, equal to 1 \cite{ulffest}. 
Its extension to nucleon-nucleon physics   
\cite{law,kolck,park,ksw96,eiras,nieves} is not 
straightforward due to the appearance of two new scales: the large 
scattering lengths of the S-waves and the large nucleon mass, $M$. The latter 
can be easily handled for B=1 although this is no longer the case for 
B$>1$ \cite{law}. Large efforts have been devoted during the last years to end with 
a convergent effective field theory (EFT) for nucleon-nucleon interactions including pions, 
for a review 
see e.g. \cite{savagefest}. On the one hand, we have the original Weinberg's proposal 
\cite{law}, with an undoubted phenomenological success \cite{epe1}. In this scheme, 
the nucleon-nucleon potential is calculated in a chiral expansion up to some definite 
order and then the Lippmann-Schwinger equations in partial waves are solved to
determine {\it numerically} the physical amplitudes. Nevertheless, 
this approach suffers from inconsistencies in the power counting due to the appearance of 
divergences that are enhanced by powers of the large nucleon mass. As a result, their 
coefficients are much larger than those expected from `naive power counting'. Despite this, 
such divergences are finally reabsorbed by counterterms whose chiral orders are 
established by assuming natural size following dimensional arguments. On the
other hand, we have the  
Kaplan-Savage-Wise (KSW) effective field theory \cite{ksw} with 
its consistent power counting that directly applies to the physical amplitudes, like in standard
CHPT \cite{wein,gl,pich}, and one finally has  $analytical$ nucleon-nucleon amplitudes. 
Interestingly, in the KSW power counting all the ultraviolet divergences appearing in loops 
are canceled by contact operators appearing at either the same or lower order in the expansion.
This is not the case in the Weinberg's approach where one has  cut-off
dependence which gives an estimate of the size of higher order corrections.
However, the convergence and the phenomenological success of the KSW effective theory 
is just restricted to a very narrow region close 
to the nucleon-nucleon threshold despite having included explicitly the pion fields 
\cite{nnlo,cohen}. 
This was clear from the analysis at NNLO in the KSW effective field theory performed 
in ref.\cite{nnlo} where the NNLO departs from data before the NLO and badly diverges 
for center-of-mass three-momentum above $\sim 100$ MeV. The reason for this bad 
convergence properties is the large contributions 
from the twice iterated pion exchange in the triplet channels. That is, pions cannot 
be treated perturbatively 
in all the nucleon-nucleon channels and in particular in the $S^3_1-D^3_1$ one.

Hence, despite all the efforts made during the past years, the issue of deriving a 
consistent effective field theory for nucleon-nucleon interactions is still open. We perform 
in this paper a step forward by extending the range of convergence of 
the KSW amplitudes. In ref. \cite{law} Weinberg proposed to calculate the 
nucleon-nucleon potential in a chiral expansion and then solve a Lippmann-Schwinger equation 
in order to treat properly the large enhancements, due to factors $2M/p$, from the reducible 
two nucleon diagrams which do not enter in the calculation of the potential, see also 
ref.\cite{ksw96}. From a S-matrix point of view the aim of the Weinberg's proposal can 
be recast so that one should resum the unitarity or right-hand cut, which is the responsible 
for all the unitarity bubbles with their large $2M/p$ factors. 
Nevertheless, solving a Lippmann-Schwinger equation is not the only way to accomplish 
this \cite{lutz} and another  one, more appropriate  
for quantum field theory  can be established. In this way, one avoids the non trivial problems 
associated with the 
renormalization of a Lippmann-Schwinger equation with highly singular potentials at the 
origin as those coming from present effective field theories. 

We propose here a general scheme to resum the unitarity cut. This scheme, 
originally motivated as an application of the N/D method \cite{chew,nd}, was  
 already 
employed in meson-meson \cite{nd,npa} and meson-baryon \cite{pin,kn} systems. 
This method should not be confused with Pad\'e resummations like those performed  
in the meson-meson or meson-baryon sectors \cite{iam}.\footnote{Later one we will show
that these Pad\'e resummations are particular cases of our formalism.} 
Nevertheless, there are specific facts in the 
nucleon-nucleon scattering, associated with the largeness of the scattering lengths in the 
S-waves, that require a special treatment not present in 
any meson-meson or meson-nucleon system,  as we explain below. In fact, a very 
reassuring feature of nucleon-nucleon 
physics is that non-perturbative effects manifests at very low center of mass 
three-momentum, $p$,  as indicated by 
the presence of bound states and poles in unphysical sheets just below threshold. 
Hence, one has at his disposal the three-momentum $p$ as a small parameter and this is 
of course at the basis of the old Effective Range Expansions (ERE)\cite{bethe}. 
It is also worth stressing that all the formalism 
presented here is analytic.

The paper is organized as follows. In sec.\ref{sec:form} the novel formalism 
contained in this work is settled. This follows the KSW power counting and the  reader unfamiliar with 
such power counting is referred to the original literature,
refs.\cite{ksw,nnlo}. It is shown how the unitarity cut can be easily resummed in terms 
of a dispersion relation of the inverse of a partial wave amplitude. As a
result an interacting kernel ${\cal R}$ arises that is determined by matching
with the pure KSW amplitudes following the KSW power counting. Once this is
done, we discuss in 
sec.\ref{sec:results} the phenomenology and compare with data. Since 
all the expressions are analytic, we fix, order by order in the expansion, many of 
the counterterms in terms of the ERE parameters $a_s$ and $r_0$ in the S-waves. 
The agreement 
with data for the S-waves and the mixing angle $S^3_1-D^3_1$ is quite remarkable in a 
broad energy range, $p\lesssim 400$ MeV ($T_{lab}
\lesssim 350$ MeV), involving all the data points of the Nijmegen partial wave 
analysis \cite{nij}. Higher partial waves are also analyzed at the same order. However, 
 only one pion exchange and the reducible part of twice iterated 
one pion exchange enter in these KSW amplitudes up to order $p$. The results
indicate that higher 
order corrections should be included so as to perform a more complete analysis 
for the $P$ and $D$ partial waves, as in the Weinberg's scheme \cite{epe1}. 
We end with some conclusions.
\section{Formalism}
\label{sec:form}
\def\theequation{\arabic{section}.\arabic{equation}}
\setcounter{equation}{0}

 It seems that pions cannot be treated perturbatively. This is indeed the 
main reason for
 the failure of convergence of the KSW scheme, which considers pions
 perturbatively in the physical amplitudes, 
 for center of mass three-momentum $p\gtrsim 100$ MeV. This was 
 clearly established in ref.\cite{nnlo}, as one can see when comparing the
calculated mixing parameter $\epsilon_1$ at NLO \cite{ksw} and NNLO 
\cite{nnlo}. While for very low momentum the NNLO calculations agree better
with data, unfortunately for momenta higher than 100 MeV the NNLO calculations 
badly diverge 
from experiment, much more than the NLO results, and no improvement is obtained
despite  pions being explicitly included. This should be compared with the
phenomenological success of the Weinberg's scheme in \cite{kolck} and 
particularly in \cite{epe1}, where the chiral expansion seems to be under 
control, at least at the level of the phenomenology. In the latter scheme there
are still the issues of avoiding cut-off dependence and the associated 
inconsistencies in the Weinberg's power counting, already referred in the 
introduction and originally established in refs.\cite{ksw96,ksw}. 

In this paper we  make use of the KSW amplitudes which are calculated within an 
effective field theory and are properly renormalized. There is nonetheless a clear
difference between the KSW effective field theory and CHPT. While the latter is convergent
in the SU(2) sector in an energy window that could be expected from the scales that are
involved in the problem,\footnote{Except in the scalar isoscalar channel.} the
same does not occur with the former \cite{nnlo}.

We want to include pions non-perturbatively but keeping at the same time the
advantages of the KSW scheme that make it a true effective field theory. In
order to accomplish this, we resum the right hand cut or unitarity cut so as to 
take care of the large $2M/p$ factors from the two nucleon reducible diagrams 
\cite{law}. This is also performed when solving a Lippmann-Schwinger
equation, as originally proposed in ref.\cite{law}, but we
will do it in a different fashion considering unitarity and analyticity in the form of a
dispersion relation of the inverse of a partial wave. 

Let us denote a partial wave by $T_{L^{2 S+1}_J,\hat{L}^{2S+1}_J}$, where $L$ and 
$\hat{L}$ refer to
the orbital angular momentum of the initial and final state respectively, 
and $S$ and $J$ indicate the total spin and total angular momentum of the
systems, in order.
Then unitarity requires above the nucleon-nucleon threshold and below the $NN\pi$ one, 
which is around $p\simeq 280$ MeV:
\ba
\label{uni1}
\hbox{Im}T_{L^{2S+1}_J,\hat{L}^{2S+1}_J}&=&\frac{M p}{4 \pi}\sum_{\ell} 
T_{L^{2S+1}_J,\ell^{2S+1}_J}
(T_{\ell^{2S+1}_J,\hat{L}^{2S+1}_J})^*~,
\ea
where $M$ is the nucleon mass. It is easy to derive from the previous equation:
\ba
\label{uni2}
\hbox{Im}\,(T^{(2S+1)J})^{-1}_{ij}=-\frac{M p}{4 \pi}\delta_{ij}~,
\ea
where $T^{(2S+1)J}$ is a matrix whose matrix elements are those triplet partial waves
that mix each other, e.g. for $S^3_1-D^3_1$ one has $T^{31}_{11}
=T_{S^3_1,S^3_1}$, $T^{31}_{12}=T^{31}_{21}=T_{S^3_1,D^3_1}$ and 
$T^{31}_{22}=T_{D^3_1,D^3_1}$. In the previous equation we denote by 
$(T^{(2S+1)J})^{-1}$ the inverse of the  $T^{(2S+1)J}$ matrix. If the partial waves
do not mix then $T^{(2S+1)J}$ is just a number  equal to 
$T_{L^{2 S+1}_J,L^{2S+1}_J}$. The imaginary part in eq.(\ref{uni2}) is responsible 
for the unitarity cut.  Thinking of a
dispersion relation for the inverse of the amplitude this cut can be easily taken 
into account from eq.(\ref{uni2}), as we previously did in the meson-meson \cite{nd} and 
meson-baryon \cite{pin,kn} sectors, and gives rise to the integral:
\ba
\label{disp}
g_i(s)=\frac{1}{\pi}\int_{4 M^2}^\infty \frac{M p(s')}{4\pi}\frac{1}{s'-s+i 0^+}\,ds' ~,
\ea
where $s$ is the usual Mandelstam variable. This integral is divergent and
requires a subtraction,  $M\nu_i/4\pi$:
\be
\label{gs}
g_i(s)=\frac{M}{4\pi}\left(\nu_i+i p+\frac{M \sigma(s)}{\pi}\log
\frac{1-\sigma(s)}{1+\sigma(s)}\right)~,
\ee
with $\sigma(s)=\sqrt{1-\frac{4 M^2}{s}}$. The logarithm is
purely real in the physical region and just gives rise to relativistic
corrections. These are obtained from eq.(\ref{disp}) but with $M p(s')/4\pi$ replaced by 
the relativistic phase space $M^2 p(s')/2\pi\sqrt{s}$ in the integrand. Since
we want to match with the KSW amplitudes, that
absorb all the relativistic corrections in the vertices,  the 
non-relativistic phase space factor $Mp/4\pi$ is used as explicit source of 
imaginary part in eq.(\ref{gs}). The relativistic corrections from the 
logarithm of eq.(\ref{gs}) are in any case essentially negligible in the
energy range that we consider. Once the unitarity cut is taken into account by
the functions $g_i(s)$,  the full partial wave matrix or number $T^{(2S+1)J}$
can be written as:
\ba
\label{keyt}
T^{(2S+1)J}=-\left[\cR^{-1}+g\right]^{-1}~.
\ea
In this expression all the other possible cuts of a nucleon-nucleon partial wave, 
either due to the exchange of other particles (in our case we have the
exchanges of pions) or because its helicity structure, are included in the
interacting kernel $\cR$ that we still must
fix. The unitarity requirements, resummation of the infinite set of reducible
diagrams with two intermediate nucleons, are accomplished by $g(s)$, which as 
stated is a diagonal matrix 
in the case of mixed partial waves, or just one function for the unmixed ones. 
For example, in the coupled partial waves $S^3_1$ and $D^3_1$, 
$g_{11}(s)=g_1(s)$ 
and corresponds to the $S^3_1$ channel and $g_{22}(s)=g_2(s)$ and refers  
to the $D^3_1$ channel. Of course, $g_{12}(s)=g_{21}(s)=0$.

We now specify $\cR$ in the key expression eq.(\ref{keyt}). For that purpose we make use of
the results of the KSW effective field theory for two nucleon systems and of its
power counting, that we apply to $\cR$ and $g(s)$ in eq.(\ref{keyt}). The
matching procedure with the KSW amplitudes can be done for any given order in the 
calculation of the KSW amplitudes and for any subtraction constants $\nu_i$ 
in $g_i(s)$, as we show explicitly below.

An important point is to establish the chiral order of $g_i(s)$. This is a trivial task for the 
phase space and the logarithmic term in $g(s)$, eq.(\ref{gs}), since they can
be expanded as a series in powers of $p$ starting at first order. The only
point we have to consider separately is the chiral 
order of $\nu_i$. For the S-waves $S^1_0$ and $S^3_1$, the corresponding
elastic KSW amplitudes start at order $p^{-1}$ and hence their inverses begin
at order $p$. Thus one can take $\nu_i$ as
 order $p$ as well, since this is the first order that appear in the
inverses of the leading KSW partial waves. Would this be the case, then $\cR$ would also
start at order $p$ and the matching with KSW amplitudes can be performed
straightforwardly. However, if we continue along these lines there is no clear 
improvement with respect to the problematic of the  KSW calculations at NNLO. 
If, on the other hand, we think of higher partial waves, e.g. $P$
and $D$, then we realize that the KSW amplitudes start at order $p^0$ so that 
 taking $g$ and $\cR$ to start as order $p^0$ is quite natural. If we
take this option as well for the S-waves, and systematically derive the $\cR$ 
matrix elements by matching with the KSW inverse amplitudes, we will see below that
the improvement with respect to the pure KSW scheme is fairly remarkable. Indeed, in
order to match with the inverses of the KSW S-wave amplitudes, that start at order $p$, we
must cancel exactly the order $p^0$ contributions in eq.(\ref{keyt}) stemming from the 
ones of $\cR$ and $g$. This fine tuning reminds of the one usually advocated to explain the
large scattering lengths in the S-waves channels \cite{law}.

In order to clarify further the previous discussion, we can easily see that considering 
$\nu_i$ as a constant of order $p^0$ is a result when $g(s)$ is calculated with a finite 
cut-off. Let us perform this illustrative exercise of more than academic 
importance since the finite three-momentum cut-off is the regularization employed in the 
Weinberg's scheme. For a given nucleon-nucleon channel $i$, we can write:
\ba
\label{cut1}
g^c_i(s)=-i 4M^2\int \frac{d^4 q}{(2\pi)^4}\frac{1}{(q^2-M^2+i 0^+)((P-q)^2-M^2+i0^+)}~,
\ea
where $P$ is the total four momentum of the two nucleon system, $P^2=s$, and
the superscript $c$ in $g^c(s)_i$ indicates that is calculated with a
three-momentum cut-off. After performing the $q^0$ integration we have:
\ba
\label{cut2}
g^c_i(s)&=&\frac{M^2}{2\pi^2}\int_M^\Lambda \frac{\sqrt{w^2-M^2}}{w^2-s/4-i0^+}\,dw~,
\ea
where $\Lambda=\sqrt{Q^2+M^2}$ with $Q$ the three-momentum cut-off. Finally we obtain the explicit result:
\ba
\label{cut3}
g^c_i(s)=\frac{M^2}{4\pi^2}\left(2\log\frac{\Lambda+Q}{M}+\sigma(s)\left[\log
\frac{\sigma(s)-\frac{2M^2+\Lambda \sqrt{s}}{Q\sqrt{s}}}{\sigma(s)+
\frac{2M^2-\Lambda\sqrt{s}}{Q\sqrt{s}}}+\log\frac{2\Lambda-\sqrt{s}}{2\Lambda+\sqrt{s}}
\right]\right)~.
\ea
Performing a non-relativistic expansion in the previous equation we have:
\ba
\label{cut4}
g^c_i(s)=\frac{M^2}{4\pi^2}\left(2\log\frac{\Lambda+Q}{M}+i\frac{\pi p}{M}+\sigma(s)
\log\frac{1-\sigma(s)}{1+\sigma(s)}+{\cal O}(\frac{p^2}{M^2})\right)~.
\ea
Comparing with $g_i(s)$, eq.(\ref{gs}), we finally have:
\be
\label{nu}
\nu_i=\frac{2M}{\pi}\log\frac{\Lambda+Q}{M}~,
\ee
which is  a quantity of order $p^0$ in the KSW power counting
\cite{ksw,nnlo}. In the KSW EFT $Q$ is 
expected to be around 300 MeV, \cite{ksw}. 
In table \ref{tab:nu} we show the values of $\nu_i$ in MeV from eq.(\ref{nu}) for different values 
of $Q$. We will obtain later, directly from fits to data, similar values of $\nu_i$.
\begin{table}
\begin{center}
\begin{tabular}{|c|c||c|c||c|c|}
\hline
$Q$ & $\nu$ & $Q$ & $\nu$ & $Q$ & $\nu$ \\
MeV & MeV & MeV & MeV & MeV & MeV\\
\hline
100 & 64 & 500 & 310 & 900 & 510 \\
200 & 130 & 600 & 360 & 1000 & 550 \\
300 & 190 & 700 & 410 & 1100 & 600 \\
400 & 250 & 800 & 460  & 1300 & 670\\
\hline
\end{tabular}
\caption{Values for $\nu_i$ from eq.(\ref{nu}) for different values of the three-momentum cut-off $Q$.
\label{tab:nu}}
\end{center}
\end{table}

Before applying the previous scheme to the KSW amplitudes, let us consider the
pedagogical example of the expansion in powers of $x$ of
$f(x)=\hbox{cot}\,x=\cos x/\sin x$. For that we write:
\ba
f(x)=-\frac{1}{\tau(x)^{-1}+\theta(x)}~,
\ea
such that $\tau(x)=t_0+t_1 x+t_2 x^2+{\cal O}(x^3)$ and $\theta(x)=z_0+z_1 x+z_2 x^2+{\cal O}(x^3)$. 
We see here that although $f(x)$ starts at order $x^{-1}$ we have considered the functions 
$\tau$ and $\theta$ to start at order $x^0$. In order to fix $t_i$ in terms of the $z_i$ (which are
assumed to be known) and of the known expansion of $\hbox{cot}\,x$, is simpler to expand the 
inverse of $f(x)$, then we obtain up to ${\cal O}(x^2)$:
\ba
\frac{\sin x}{\cos x}=x+{\cal O}(x^3)=-\frac{1}{t_0}+\frac{t_1}{t_0^2} x+
\frac{t_2}{t_0^2} x^2-\frac{t^2_1}{t_0^3} x^2-z_0-z_1 x-z_2 x^2+{\cal O}(x^3)~.
\ea
It follows then:
\ba
t_0&=&-\frac{1}{z_0}\nn\\
t_1&=&\frac{(z_1+1)}{z_0^2}\nn\\
t_2&=&\frac{z_2}{z_0^2}+\frac{(z_1+1)^2}{z_0^3}~.
\ea
It is obvious how to proceed for higher orders. This simple example also illustrates that if we 
want  to
calculate $\tau(x)$ up to order $x^i$ one needs to know $f(x)$ up to order $x^{i-2}$ since $f(x)$ already
starts at order $x^{-1}$.

Let us now analyze carefully the elastic $S^1_0$ channel following the previous 
scheme. After that we will present more briefly the analogous procedure in the 
$S^3_1-D^3_1$ coupled channel sector and for the $P$, $D$, $F^2_2$ and $G^3_3$ waves. 

\subsection{$S^1_0$ elastic partial wave}

The KSW $S^1_0$ partial wave, $A^{KSW}_{S^1_0}$, was calculated at NLO 
(order $p^0$) in ref.\cite{ksw}
and then at NNLO (order $p$) in ref.\cite{nnlo}. Let us denote these partial
waves by $A_{-1}(p)$, $A_0(p)$ and $A_1(p)$, where the subscript indicates the
KSW order. As in the simple example of the $\hbox{cot}\,x$, if we take
as input the KSW amplitudes up to order $p$ then we will be able to calculate $\cR$ up 
to order $p^3$. This unambiguously fixes the order one has to calculate in the KSW 
EFT so that $\cR$ is obtained up to the required precision. Following the same notation 
as for the KSW amplitudes, let us  write:
\be
\cR=R_0+R_1+R_2+R_3+\Opc~,
\ee 
and $g(s)=M\nu/4\pi+iM p/4\pi-p^2/2\pi^2+\Opc$. Then we
we must match:
\ba
\label{kswexp}
\frac{1}{A^{KSW}_{S^1_0}}=\left(\frac{1}{A_{-1}}\right)-
\left(\frac{A_0}{A_{-1}^2}\right)
+\left(\frac{A_0^2-A_1 A_{-1}}{A_{-1}^3}\right)+\Opc~,
\ea
with
\ba
\label{mineexp}
\!-\!\left(\frac{1}{\cR}+g\right)\!\!\!\!&=&\!\!\!\!
\!-\!\left(\frac{1}{R_0}+\frac{M\nu}{4\pi}\right)
\!+\!\left(\frac{R_1}{R_0^2}-i\frac{Mp}{4\pi}\right)
\!+\!\left(\frac{p^2}{2\pi^2}+\frac{R_0 R_2-R_1^2}{R_0^3}\right)
\!+\!\left(\frac{R_1^3-2R_0 R_1 R_2+R_0^2R_3}{R_0^4}\right) \nn \\
&+&\Opc~,
\ea
where  we have shown between brackets the different orders, from order $p^0$ up to 
order $p^3$. As a result of the matching we can fix $R_0$, $R_1$,
$R_2$ and $R_3$ in terms of $\nu$, $A_{-1}(p)$, $A_0(p)$ and $A_1(p)$. We follow 
ref.\cite{nnlo}  for expressing the KSW amplitudes, so that any of the 
previous amplitudes are scale  independent at each order in the expansion.

Taking into account that 
\be
A_{-1}=-\frac{4\pi}{M}\frac{1}{\gamma+ip}~,
\ee
with $\gamma$ a quantity of order $p$, we can write the following expressions for the
$R_i$:
\ba
\label{exp}
R_0&=&-\frac{4\pi}{M\nu}~,\nn\\
R_1&=&-\frac{4\gamma \pi}{M\nu^2}~,\nn\\
R_2&=&-\frac{4(2\nu p^2+\gamma^2 M\pi)
+\nu(\frac{4\pi}{A_{-1}})^2 A_0(p)}{M^2\nu^3}~,\nn\\
R_3&=&-\frac{1}{4 M^2 \nu^4\pi}(8\gamma  \nu(\frac{4\pi}{A_{-1}})^2\pi 
A_0(p) 
-\nu^2(\frac{4\pi}{A_{-1}})^3 A_0(p)^2+4\pi[4\gamma(4\nu p^2+
\gamma^2 M \pi)
\nn\\
&+& 
\nu^2(\frac{4\pi}{A_{-1}})^2 A_1(p)])~.
\ea
Hence working at NLO, ${\cal O}(p^0)$, in the KSW amplitudes \cite{ksw} we will have $\cR$ up 
to ${\cal O}(p^2)$, as explained above,
\be
\label{rnlo1s0}
\cR^{NLO}=R_0+R_1+R_2~,
\ee
and matching with the ones at NNLO \cite{nnlo}, ${\cal O}(p)$, we calculate $\cR$ up to 
${\cal O}(p^3)$:
\be
\label{rnnlo1s0}
\cR^{NNLO}=R_0+R_1+R_2+R_3~.
\ee
The resulting $\cR$ is substituted in eq.(\ref{keyt}) and in this way we calculate 
the partial waves at the different orders considered so forth. It is clear that
the process above can be done so as to match with a KSW amplitude calculated at 
any order. In this way the precision of the resulting amplitude is increased
order by order.

        Let us consider now in more detail the meaning of the kernel ${\cal R}$. 
Take first the simpler case of the pionless effective field theory, where only local 
interactions and even numbers of derivatives appear in the Lagrangians. For
this case, the resulting partial waves have only the right hand cut and are
free of crossed cuts due to the exchanges of other particles, being the pions
the lightest ones. This is important for the nucleon-nucleon dynamics 
since at very low energies the pions can be treated as heavy particles and integrated 
out \cite{nopions}. Thus, only local operators and the unitarity cut remain. 
We follow the line of reasoning of
ref.\cite{castillejo}, where the so important Castillejo-Dalitz-Dyson (CDD) poles
were introduced. This reference was also used in ref.\cite{nd} in the context
of chiral Lagrangians to show the general structure of a meson-meson partial
wave with only the unitarity cut. Here we show the main points of the
reasoning, for further details the reader is referred to ref.\cite{nd}. The
idea is that when the partial wave $T_{L^{2S+1}_J,L^{2S+1}_J}$ has a zero then
its inverse has a pole. This pole in the inverse of the partial wave can be on
the real axis or in other place of the complex plane of the physical sheet
($\hbox{Im}p\geq 0$). Special care is needed when these poles lie on the real 
axis above threshold since then eq.(\ref{uni1}) can only be applied between 
any pair of such poles. These technical details are given in 
refs.\cite{castillejo,nd}. In this way, the inverse of the partial wave is a
meromorphic function in the cut $s$ plane from threshold to infinity. The net answer
from a dispersion relation of the inverse of the amplitude, by applying the
Cauchy theorem of integration to the circuit made up by the circle at infinity
deformed to engulf the unitarity cut, is:
\be
T_{L^{2S+1}_J,L^{2S+1}_J}=-\left(\sum_n \frac{\gamma_n}{s-s_n}+g\right)^{-1}~,
\label{cdd}
\ee
where the poles present in the sum are the so called CDD poles \cite{castillejo}. 
The sum is what we have denoted before in eq.(\ref{keyt}) by $1/{\cal R}$. 
Let us note that in the pionless 
effective field theory supplied with the PDS scheme, the Schr\"odinger equation can 
be solved straightforwardly and results \cite{ksw}:
\be
\label{kswsch}
A=-\left(\frac{1}{\sum_{m=0} C_{2m} p^{2m}}+\frac{M}{4\pi}(\mu+ip) \right)^{-1}~,
\ee
with $A$ being the same partial wave as $T_{L^{2S+1}_J,L^{2S+1}_J}$. 
Comparing this equation with eq.(\ref{cdd}), one has:
\be
\frac{1}{{\cal R}}=\sum_n \frac{\gamma_n}{s-s_n}=\frac{1}{\sum_{m=0}C_{2m} p^{2m}}+\frac{M}{4\pi}(\mu-\nu)=\frac{1}{\sum_{m=0}C'_m p^{2m}}~,
\label{local}
\ee
where we have omitted in the previous equation the relativistic corrections in $g$ 
from the logarithmic term in eq.(\ref{gs}). The $C'_m$ are given in terms of the $C_m$ so as to reabsorb the constant term in the sum. One can always choose the $s_n$ and $\gamma_n$ so that the previous equality holds.\footnote{For example, prove the above equality by considering first that the sum on the right hand side just contains two terms. Then, for the general case, it is easily seen, by recurrence to the case with one term less in the sum, that the equality holds. Note that when $s_n\rightarrow \infty$ the corresponding CDD pole just gives rise to a constant}. This is why we have written 
finally $1/{\cal R}$ in eq.(\ref{keyt}), instead of simply ${\cal
  R}$, because at the tree level, omitting the $g$ function in eq.(\ref{cdd}), one 
has $T_{L^{2S+1}_J,L^{2S+1}_J}={\cal R}$. Let us note in addition that 
$-{\cal R}$ can be identified with the renormalized potential if we rewrite 
$T=-1/(1/{\cal R}+g)$ as:
\be
\label{lse}
T=-{\cal R}-{\cal R}\,g\,T~,
\ee
like an ordinary Lippmann-Schwinger equation with an on-shell potential
$V\equiv -{\cal R}$. Let us remind, as stated above, that $g$ is just the 
unitarity bubble iterated when solving a Lippmann-Schwinger equation. 

When the pions are included in the formalism, the
interacting kernel ${\cal R}$ contains apart from a {\it finite} sum of local
terms, like those of eq.(\ref{local}), other contributions coming from the exchange 
of pions which give rise to crossed cuts. All these contributions are included 
{\it perturbatively} when
${\cal R}$ is fixed by matching the expansion of the inverse of eq.(\ref{keyt}) with 
the expansions  of the inverses of the KSW partial waves.  Let us stress that the entire 
formalism is algebraic since the interacting  kernel 
${\cal R}$ is on-shell. This notorious
simplicity of our approach compared to that of refs.\cite{law,kolck,epe1} is a result of
resumming the unitarity bubbles by making use of analyticity and unitarity that
only involve on-shell amplitudes. One needs to realize that for resumming 
the large factors $2M/p$ associated with two-nucleon intermediate
states, whose necessity was stressed in ref.\cite{law}, there are several
possibilities. One way is to solve
the Lippmann-Schwinger equation \cite{law,kolck,epe1}, but  there is
another standard method based on the separation of cuts resulting from
S-matrix theory, this is the N/D method \cite{chew} that we follow here.

\subsection{$S^3_1$--$D^3_1$ coupled partial waves}

The resulting expressions to match between are the expansions of the matrix
elements of the
inverse of the KSW matrix of partial waves for the $S^3_1-D^3_1$ sector, 
 $A^{KSW}_{S^3_1-D^3_1}$:
\ba
(A^{KSW}_{S^3_1-D^3_1})^{-1}_{11}&=&\left(\frac{1}{A_{11,-1}}\right)+
\left(\frac{A_{12,0}^2-A_{11,0}A_{22,0}}{A_{11,-1}^2 A_{22,0}}\right)
\nn\\
&+&
\left(\frac{A_{12,0}^4+2A_{11,-1} A_{12,0}A_{12,1}A_{22,0}+
(A_{11,0}^2-A_{11,-1}A_{11,1})A_{22,0}^2}{A_{11,-1}^3 A_{22,0}^2}
\right.\nn\\
&-&\left.\frac{A_{12,0}^2(2A_{11,0}A_{22,0} +
A_{11,-1}A_{22,1})}{A_{11,-1}^3 A_{22,0}^2}\right)+\Opc~,\nn\\
(A^{KSW}_{S^3_1-D^3_1})^{-1}_{12}&=&-\left(
\frac{A_{12,0}}{A_{11,-1}A_{22,0}}\right)-\left(\frac{A_{12,0}^3+
A_{11,-1} A_{12,1}A_{22,0}-A_{12,0}(A_{11,0}A_{22,0}+A_{11,-1}
A_{22,1})}{A_{11,-1}^2 A_{22,0}^2}\right)\nn\\&+&\Opt~,\nn\\
(A^{KSW}_{S^3_1-D^3_1})^{-1}_{22}&=&\left(\frac{1}{A_{22,0}}\right)
+\left(\frac{A_{12,0}^2-A_{11,-1}A_{22,1}}{A_{11,-1}A_{22,0}^2}\right)
+\Opd~,
\label{ksw3sdexp}
\ea
and the ones that result from the expansion of eq.(\ref{keyt}). In this case 
$\cR$ is a $2\times 2$ symmetric matrix:
\ba
\cR=\left(
\begin{array}{ll}
R_{11} & R_{12}\\
R_{12} & R_{22}
\end{array}\right)~,
\ea 
and $g(s)=diagonal(g_1(s),g_2(s))$ with its associated subtraction constants 
$\nu_1$ and $\nu_2$. Like in the 
$S^1_0$ channel we will take $g_i={\cal O}(p^0)$ and $R_{11}=R_{11,0}+
R_{11,1}+R_{11,2}+R_{11,3}+\Opc$. In the KSW scheme at the leading order the
$S^3_1$ and $D_1^3$ are uncoupled and the mixing starts at NLO, one order higher. 
Thus, we take $R_{12}=R_{12,1}+R_{12,2}+\Opt$, starting one order higher than $R_{11}$. 
Finally, since the $D^3_1$ partial wave starts at order $p^0$, and is free of 
unnatural scattering lengths, we take then $R_{22}=R_{22,0}+R_{22,1}+\Opd$. 
The expansion of the inverse of eq.(\ref{keyt}) is now straightforward and one obtains: 
\ba
\label{expansionT}
(T^{31})^{-1}_{11}&=&-\!\left(\frac{M\nu_1}{4\pi}+\frac{1}{R_{11,0}}\right)
\!+\!\left(\frac{R_{11,1}}{R_{11,0}^2}-i\frac{Mp}{4\pi}\right)
\!+\!\left(\frac{p^2}{2 \pi^2}+\frac{R_{11,0}R_{11,2}R_{22,0}-R_{11,1}^2 
R_{22,0}-R_{11,0}R_{12,1}^2}{R_{11,0}^3 R_{22,0}}\right)\nn\\
&-&\left(\frac{-R_{11,1}^3 R_{22,0}^2+
2R_{11,0}R_{11,1}R_{22,0}(-R_{12,1}^2+R_{11,2}R_{22,0})}{R_{11,0}^4 R_{22,0}^2}
\right.\nn\\
&-&\left.\frac{R_{11,0}^2(-2R_{12,1}R_{12,2}R_{22,0}+R_{11,3}R_{22,0}^2
+R_{12,1}^2R_{22,1})}{R_{11,0}^4 R_{22,0}^2} \right)+\Opc~,\nn\\
(T^{31})^{-1}_{12}&=&\left(\frac{R_{12,1}}{R_{11,0}R_{22,0}}\right)
-\left(\frac{R_{11,1}R_{12,1}R_{22,0}-R_{11,0}R_{12,2}R_{22,0}+R_{11,0}
R_{12,1}R_{22,1}}{R_{11,0}^2R_{22,0}^2}\right)+\Opt\nn~,\\
(T^{31})^{-1}_{22}&=&-\left(\frac{M\nu_2}{4\pi}+\frac{1}{R_{22,0}}\right)
+\left(
\frac{R_{22,1}}{R_{22,0}^2}-i\frac{M p}{4\pi}\right)+\Opd~.
\ea
We can then easily solve  for the $R_{ij,k}$ and we obtain:
\ba
\label{r3s1}
R_{11,0}&=&-\frac{4\pi}{M\nu_1}~,\nn\\
R_{11,1}&=&-\frac{4\gamma\pi}{M\nu_1^2}~,\nn\\
R_{11,2}&=&\frac{1}{M^2\nu_1^3(4\pi+M\nu_2A_{22,0})}
\left\{M^3\nu_1\nu_2(\gamma+ip)^2A_{12,0}^2-4(2\nu_1p^2+\gamma^2M\pi)
(4\pi+M\nu_2A_{22,0})\right.\nn\\
&-&\left.M^2\nu_1(\gamma+ip)^2A_{11,0}(4\pi+M\nu_2A_{22,0})
\right\}~,\nn\\
R_{11,3}&=&\frac{1}{4M\nu_1^2\pi A_{22,0}^2}\left\{
-M(\gamma+ip)^2(M(\gamma+ip)A_{12,0}^4-8\pi A_{12,0}A_{12,1}A_{22,0}
+(M(\gamma+ip)A_{11,0}^2\right.
\nn\\
&+&\left.
4\pi A_{11,1})A_{22,0}^2-
2A_{12,0}^2(M(\gamma+ip)A_{11,0}A_{22,0}-2\pi A_{22,1}))
+\frac{1}{\nu_1^2(4\pi+M\nu_2 A_{22,0})^2}
\left(
\right.\right.\nn\\
&\times&\left.\left.
8\pi(2\gamma^3\pi A_{22,0}^2(4\pi+M\nu_2 A_{22,0})^2-
\frac{1}{M}(\gamma A_{22,0}(4\pi+M\nu_2 A_{22,0})^2(-M^2\nu_1(\gamma+ip)^2
A_{12,0}^2\right.\right.\nn\\
&+&\left.\left.(4(2\nu_1 p^2+\gamma^2 M\pi)+M^2\nu_1(\gamma+ip)^2 
A_{11,0})A_{22,0}))-M\nu_1(\gamma+ip)^2A_{12,0}(-M\nu_1(\gamma+ip)
A_{12,0}^3\right.\right.\nn\\
&\times&\left.\left.
(2\pi+M\nu_2 A_{22,0})+4\nu_1\pi A_{12,1}A_{22,0}(4\pi
+M\nu_2 A_{22,0})+A_{12,0}(M(4\gamma \nu_2 \pi-2i\nu_1 p\pi 
\right.\right.\nn\\
&+&\left.\left.
M\nu_1\nu_2(\gamma+ip)A_{11,0})A_{22,0}^2-8\nu_1\pi^2 A_{22,1}+
4\pi A_{22,0}(4\gamma \pi+M\nu_1(\gamma+ip)A_{11,0}
\right.\right.\nn\\
&-&\left.\left.M\nu_1\nu_2 A_{22,1}))))\right)\right\}~,\nn\\
R_{12,1}&=&\frac{4(\gamma+ip)\pi A_{12,0}}{\nu_1(4\pi+M\nu_2 A_{22,0})}~,\nn\\
R_{12,2}&=&\frac{1}{\nu_1^2(4\pi+M\nu_2 A_{22,0})^2}
\left\{(\gamma+ip)(-M^2\nu_1\nu_2(\gamma+ip)A_{12,0}^3+4\nu_1 \pi 
A_{12,1}(4\pi+M\nu_2A_{22,0})\right.\nn\\
&+&\left.A_{12,0}(M\nu_1(\gamma+ip)A_{11,0}
(4\pi+M\nu_2 A_{22,0})+4\pi(4\gamma\pi+M(\gamma\nu_2-i\nu_1 p)A_{22,0}-
M\nu_1\nu_2 A_{22,1})))\right\}~,\nn\\
R_{22,0}&=&-\frac{4\pi A_{22,0}}{4\pi+M\nu_2 A_{22,0}}~,\nn\\
R_{22,1}&=&-\frac{4\pi(M(\gamma+ip)A_{12,0}^2-iM p A_{22,0}^2+4\pi A_{22,1})}
{(4\pi+M\nu_2 A_{22,0})^2}~.
\ea
Similarly as in the $S^1_0$ case, working at NLO implies:
\ba
R^{NLO}_{11}&=&R_{11,0}+R_{11,1}+R_{11,2}~,\nn\\
R^{NLO}_{12}&=&R_{12,1}~,\nn\\
R^{NLO}_{22}&=&R_{22,0}~,
\ea
and at NNLO:
\ba
R^{NNLO}_{11}&=&R_{11,0}+R_{11,1}+R_{11,2}+R_{11,3}~,\nn\\
R^{NNLO}_{12}&=&R_{12,1}+R_{12,2}~,\nn\\
R^{NNLO}_{22}&=&R_{22,0}+R_{22,1}~.
\ea

\subsection{$P$, $D$, $F^3_2$ and $G^3_3$ partial waves except $D^3_1$}

For the $P$ and $D$ waves, the NLO KSW amplitudes 
\cite{ksw} just contain one pion exchange (OPE) and at NNLO 
\cite{nnlo,kaiser} they only include in addition the reducible part of the twice 
iterated  OPE. The physics behind this is then quite 
limited and will show up in the phenomenology which, on the other hand, is of the 
same quality as that of the LO Weinberg's scheme results, 
ref.\cite{epe1}. Indeed, at LO the potential within the Weinberg's scheme just
contains OPE (our order $p^0$ contribution) while the
reducible part of the twice iterated OPE is generated by  solving the
Schr\"odinger equation. For these partial waves, which are free from
enhancements due to unnatural scattering lengths, the formalism simplifies
since the scaling of the counterterms is just given by dimensional analysis and
there are no KSW amplitudes of order $p^{-1}$. For the elastic ones, $\cR=R_0+R_1+\Opd$ 
and $g={\cal O}(p^0)$ and for the coupled channel
partial waves, namely $P^3_2-F^3_2$ and $D^3_3-G^3_3$,
$R_{ij}=R_{ij,0}+R_{ij,1}+\Opd$ and $g_i(s)={\cal O}(p^0)$, 
since all the channels start to contribute at the same order $p^0$ and as
stated they are free of unnatural scattering lengths.

For the partial waves without coupled channels, after performing the appropriate 
KSW expansion and matching with the inverse of the KSW amplitude, as done previously, 
one can write:
\ba
R_{0}&=&-\frac{A_0}{1+A_0\frac{M\nu}{4\pi}}~,\nn\\
R_{1}&=&-\frac{1}{(1+A_0\frac{M\nu}{4\pi})^2}(A_1-i\frac{p M}{4\pi}A_0^2)~,
\label{expnatu}
\ea
so that,
\ba
\cR^{NLO}&=&R_0~,\nn\\
\cR^{NNLO}&=&R_0+R_1 ~.
\ea
For the two coupled channel partial waves, the chiral expansion is performed
in coupled channels. Taking into account that $R_{ij}=R_{ij,0}+R_{ij,1}+\Opd$, 
as discussed, one has:
\ba
R_{11,0}&=&-\frac{4\pi(4\pi A_{11,0}-M \nu_2 A_{12,0}^2+M\nu_2 A_{11,0}
A_{22,0})}{(4\pi+M \nu_1 A_{11,0})(4\pi+M\nu_2 A_{22,0})-M^2
 \nu_1\nu_2A_{12,0}^2}~,
\nn\\
R_{11,1}&=&\frac{1}
{\left[M^2\nu_1\nu_2A_{12,0}^2-(4\pi+M\nu_1A_{11,0})
(4\pi+M\nu_2A_{22,0})\right]^2}\left\{
4i\pi(4i\pi A_{11,1}(4\pi+M\nu_2A_{22,0})^2 \right.\nn\\
&+&\left.
M(M^2\nu_2^2 p A_{12,0}^4-8i\nu_2\pi A_{12,1}A_{12,0}(4\pi+M\nu_2A_{22,0})+
pA_{11,0}^2(4\pi+M\nu_2A_{22,0})^2\right.\nn\\
&-&\left.
2A_{12,0}^2(-2iM\nu_2^2\pi A_{22,1}+
p(4\pi(-2\pi+M\nu_2A_{11,0})+M^2\nu_2^2A_{11,0}A_{22,0}))))\right\}~,\nn\\
R_{12,0}&=&-\frac{16\pi^2A_{12,0}}
{(4\pi+M \nu_1 A_{11,0})(4\pi+M\nu_2 A_{22,0})-M^2 \nu_1\nu_2A_{12,0}^2}
~,\nn\\
R_{12,1}&=&-\frac{1}{\left[M^2\nu_1\nu_2A_{12,0}^2-(4\pi+M\nu_1A_{11,0})
(4\pi+M\nu_2A_{22,0})\right]^2}\left\{
16\pi^2(A_{12,1}(M^2\nu_1\nu_2A_{12,0}^2\right.\nn\\
&+&\left.(4\pi+M\nu_1
A_{11,0})(4\pi+M\nu_2A_{22,0}))+iMA_{12,0}(-4p\pi A_{11,0}+M(\nu_1+\nu_2)p
A_{12,0}^2+4i\nu_2\pi A_{22,1}\right.\nn\\
&+&\left.
iM\nu_1\nu_2A_{11,0}A_{22,1}-4p\pi A_{22,0}
-M\nu_1pA_{11,0}A_{22,0}-M\nu_2pA_{11,0}A_{22,0}\right.\nn\\
&+&\left.i\nu_1A_{22,1}
(4\pi+M\nu_2A_{22,0})))\right\}~,\nn\\
R_{22,0}&=&-\frac{4\pi(-M \nu_1 A_{12,0}^2+(4\pi+M\nu_1 A_{11,0})A_{22,0})}
{(4\pi+M \nu_1 A_{11,0})(4\pi+M\nu_2 A_{22,0})-M^2 \nu_1\nu_2A_{12,0}^2}~,\nn\\
R_{22,1}&=&\frac{1}{\left[M^2\nu_1\nu_2A_{12,0}^2\!-\!(4\pi+M\nu_1A_{11,0})
(4\pi+M\nu_2A_{22,0})\right]^2}\left\{
4i\pi(\!-8iM\nu_1\pi(4\pi\!+\!M\nu_1A_{11,0})A_{12,1}A_{12,0}
\right.\nn\\
&+&\left.
M^3\nu_1^2pA_{12,0}^4+(4\pi+M\nu_1A_{11,0})^2(4i\pi A_{22,1}+MpA_{22,0}^2)-
2MA_{12,0}^2(-2iM\nu_1^2\pi A_{11,1}\right.\nn\\
&+&\left.p(-8\pi^2+M\nu_1(4\pi+M\nu_1A_{11,0})
A_{22,0})))\right\}~,
\ea
where the subscript $1$ before the period always refers to the channel with 
lower orbital angular momentum and $2$ to the highest one. 

Working at NLO implies taking:
\ba
R^{NLO}_{11}&=&R_{11,0}~,\nn\\
R^{NLO}_{12}&=&R_{12,0}~,\nn\\
R^{NLO}_{22}&=&R_{22,0}~,
\ea
and at NNLO:
\ba
R^{NNLO}_{11}&=&R_{11,0}+R_{11,1}~,\nn\\
R^{NNLO}_{12}&=&R_{12,0}+R_{12,1}~,\nn\\
R^{NNLO}_{22}&=&R_{22,0}+R_{22,1}~.
\ea

As a check for all the previous expressions for the interacting kernel ${\cal
  R}$, we have explicitly verified that they are real 
in the physical nucleon-nucleon region, $p$ real and positive, and furthermore 
they also have the correct KSW order, as they should.

\section{Results and discussion}
\label{sec:results}
\def\theequation{\arabic{section}.\arabic{equation}}
\setcounter{equation}{0}

In this section we  consider the phenomenological applications of the
previous scheme to the S and higher partial waves. Of particular relevance is
to study
the triplet S-wave channel, $S^3_1$, and its mixing with the $D^3_1$, since here 
the KSW amplitudes
do not converge for $p\gtrsim 100$ MeV although pions are explicitly included.
We first discuss the $S^1_0$ channel and then consider the $S^3_1$ coupled with
the $D^3_1$ partial wave. After that we turn to discuss the $P$, $D$, $F^3_2$ and 
$G^3_3$  waves and compare with other approaches.

\subsection{ $S^1_0$ channel}

We follow the notation of ref.\cite{nnlo} for the KSW amplitudes, where only 
those combinations of
counterterms that appear in a given amplitude are shown and are denoted by 
$\xi_i$. At NLO we have two of such counterterms, $\xi_1$ and $\xi_2$ together 
with $\gamma$ that already appears at LO. We express these  two
counterterms in terms of $\gamma$ by performing the ERE in eq.(\ref{keyt}), 
with $\cR^{NLO}$  given in eq.(\ref{rnlo1s0}), reproducing the physical values
 of the scattering length, $a_s$, and effective range, $r_0$. We then have:
\ba
\label{z1nlo1s0}
\xi_1&=&\frac{-g_A^2M^2(6\gamma^2-8\gamma m_\pi+3m_\pi^2)(-1+a_s\nu)^2+
12f^2 m_\pi^2(4-8a_s\nu+a_s^2 M \nu^2\pi r_0)}{96 f^2 m_\pi^2(-1+a_s\nu)^2\pi^2}
~,\nn\\
\xi_2&=&\frac{M(g_A^2 \gamma M(\gamma-2 m_\pi)\nu(-1+a_s\nu)+8f^2(-\nu^2 
+\gamma^2(-1+a_s\nu)+\gamma \nu (-1+a_s\nu))\pi)}{32 f^2 m_\pi^2 
\nu(-1+a_s\nu)\pi^2}~.
\ea
At NNLO there are three more counterterms, $\xi_3$, $\xi_4$ and $\xi_5$. Once
the renormalization group equations are solved within the KSW perturbative
scheme, $\xi_5$ results to be a higher order counterterm and must be set
equal to zero at NNLO. We will show below, when
discussing the results for the $S^3_1-D^3_1$ channel after eq.(\ref{xi5}), 
that indeed, $\xi_5$ must be left as a free parameter within our scheme. 
Nevertheless, we found after performing fits with $\xi_5$ free,
that it turns out to be negligibly small in any case, so that 
in the following we also make $\xi_5=0$ for this channel as in the pure KSW
treatment \cite{nnlo}. We then fix the counterterms 
$\xi_3$ and $\xi_4$ in terms of $\xi_1$, $\xi_2$, $\gamma$, 
$a_s$ and $r_0$ by performing the ERE. The expressions of $\xi_3$ and 
$\xi_4$ in terms of $\xi_1$, $\xi_2$, $\gamma$, $a_s$ and $r_0$ are:
\ba
\label{z3nnlo1s0}
\xi_3&=&\frac{1}{1536 f^4 m_\pi^2\nu^2(-1+a_s\nu)\pi^3}
\left\{48f^2g_A^2M\nu(-1+a_s\nu)\pi(2\gamma^3M+\gamma^2 M
(-4 m_\pi+\nu)+8m_\pi^3\nu \pi \xi_2\right.\nn\\
&-&\left.2\gamma m_\pi \nu(M+4m_\pi \pi 
\xi_2))-384 f^4\pi^2(\gamma^2 M\nu(1-a_s\nu)+\gamma^3(M-a_sM\nu)-
\gamma \nu(-1+a_s\nu)\right.\nn\\
&\times&\left.(M\nu-8m_\pi^2 \pi \xi_2)+\nu^2(M\nu+4m_\pi^2
(-1+a_s\nu)\pi \xi_2))+g_A^4 M^3\nu^2(-1+a_s\nu)(6\gamma^3
-21\gamma^2 m_\pi\right.\nn\\
&-&\left.
\gamma m_\pi^2(-18+\log(4096))+m_\pi^3 \log(4096))\right\}~,
\nn\\
\xi_4&=&-\frac{1}{3072 f^4 m_\pi^2\nu(-1+a_s\nu)^2\pi^3}
\left\{384 f^4 m_\pi^2\pi(8\gamma(-1+a_s\nu)^2(-1+2\pi^2\xi_1)+
\nu(-4+8\pi^2\xi_1\right.\nn\\
&+&\left.a_s^2\nu^2\pi(-Mr_0+8\pi\xi_1)+8
a_s\nu(1-2\pi^2\xi_1)))+32f^2g_A^2M(-1+a_s\nu)^2\pi(12\gamma^3M+
\gamma^2M(-16 m_\pi+6\nu)\right.\nn\\
&+&\left.2\gamma m_\pi(M(3m_\pi-4\nu)+12m_\pi\nu\pi
(\xi_1-2\xi_2))+m_\pi^2\nu(3M+8m_\pi\pi(-3\xi_1+4\xi_2)))\right.\nn\\
&+&\left.
g_A^4M^3\nu(-1+a_s\nu)^2(48\gamma^3-135 \gamma^2m_\pi+2m_\pi^3
(-13+4\log(16))-4\gamma m_\pi^2(-27+\log(4096)))\right\}~.
\ea
The values for the parameters that we take are 
$f=130.67$ MeV, $g_A=1.267$, $m_\pi=138$ MeV, $M=939$ MeV. Specifically for
the $S^1_0$ channel, the ERE parameters are $a_s=-23.714$ fm and 
$r_0=2.73$ fm. On the other hand, since $a_s$ is so large in this channel, 
and $\gamma$ is around $1/a_s$ both in KSW as in our approach,\footnote{For example, 
just take ${\cal  R}^{LO}=R_0+R_1$ in eq.(\ref{keyt}) and then perform the
ERE, with $R_0$ and $R_1$ given in eq.(\ref{exp}).} in the following we take 
directly $\gamma=0$.

\begin{figure}[H]
\psfrag{1S0(degrees)}{$S^1_0$; $\delta$(deg)}
\psfrag{p_cm(MeV)}{p (MeV)}
\centerline{\epsfig{file=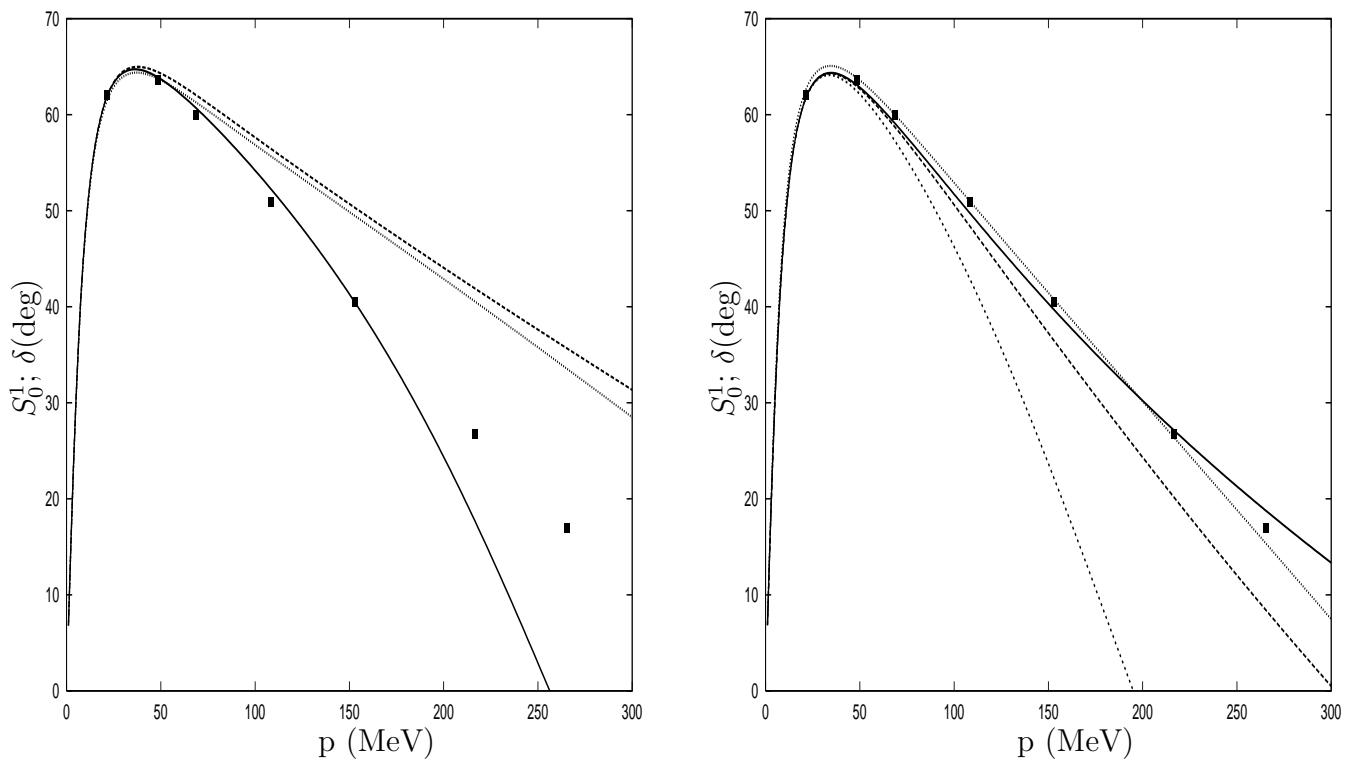,height=4.0in,width=7.0in,angle=0}}
\vspace{0.2cm}
\caption[pilf]{\protect \small
 Phase shifts for the $S^1_0$ channel. The
 dotted lines  in the left and right panels are the NLO/NNLO KSW results \cite{nnlo}, 
respectively. Left panel: The solid and dashed lines represent the NLO results
 of our approach with $\nu=200$ MeV and 900 MeV, respectively. Right panel:
 The solid line is the fit of eq.(\ref{solonu}). The dashed and short-dashed
 lines are our NNLO results with $\nu=500$ MeV and 200 MeV, in order. The data 
are from the Nijmegen partial wave analysis, ref.\cite{nij}.
\label{fig:1s0kswcm}}
\end{figure}

In the KSW approach at NLO one has the free parameters $\gamma$, $\xi_1$ and
$\xi_2$. In ref.\cite{nnlo} $\gamma$ is fixed by requiring the presence of a 
pole in the unphysical sheet in the position required by the ERE, just below 
threshold. Then $\xi_2$ is fixed in terms of $\xi_1$ to avoid spurious poles
that appear at NLO (double poles). As a result only $\xi_1$ is taken as free
in the fit to the elastic $S^1_0$ phase shifts in ref.\cite{nnlo}. The 
resulting values are:
\be
\begin{array}{lll}
\gamma^\star=-7.88\,\,\hbox{fm}^{-1}~,& \xi_1=0.22~,& \xi_2^\star=0.03~,
\end{array}
\label{1s0nloksw}
\ee 
where those parameters marked with a star are not taken as free ones in
the fit. The corresponding curve is the  dotted one in the left panel of 
fig.\ref{fig:1s0kswcm}. In the approach that we present here we have one
more parameter at NLO, $\nu$. Since, as discussed above, we have fixed $\gamma=0$
and $\xi_1$ and $\xi_2$ are calculated from the ERE, eq.(\ref{z1nlo1s0}), only
$\nu$ would be free in the fit. Instead of performing a fit, and in order to show as
well the sensitivity of our results at NLO under changes of the subtraction 
constant $\nu$, we show in the left panel of fig.\ref{fig:1s0kswcm} two curves 
corresponding to $\nu=200$ (solid line) and $\nu=900$ MeV (dashed line). 
The solid curve reproduces already very well the data for $p\lesssim 150$
MeV. The calculated counterterms are $\xi^\star_1=0.22(0.25)$,
$\xi^\star_2=0.03(0.03)$ for $\nu=200(900)$. 
Taking into account table \ref{tab:nu} we see that the considered variation of
$\nu$ implies a large change of the hypothetical cut-off from around 300 MeV to
values around 2 GeV. It is also worth mentioning the constancy of the values of
the counterterms $\xi_1$ and $\xi_2$. This is just a consequence of the fact 
that $\nu a_s>>1$ as can be seen from eq.(\ref{z1nlo1s0}).

Let us consider now the NNLO results both from the pure KSW approach and
ours. At this order three new counterterms appear, $\xi_3$, $\xi_4$ and $\xi_5$. 
Because $\xi_5$ is a higher order counterterm is fixed at zero in
ref.\cite{nnlo}. On the other hand, by imposing, as in the NLO case, the absence 
of double and triple poles one can express $\xi_3$ in terms of $\xi_4$. As a 
result only two
free parameters are left to fit the data. The resulting values from 
ref.\cite{nnlo} are:
\be
\begin{array}{lllll}
\gamma^\star=-7.88\,\,\hbox{fm}^{-1}~,& \xi_1=0.078~,& \xi_2^\star=0.03~, &
\xi_3^\star=0.18~, & \xi_4=0.25~.
\end{array}
\label{1s0nnloksw}
\ee 
The generated curve is the dotted one in the right panel of
fig.\ref{fig:1s0kswcm} which indeed reproduces the data rather accurately. We
now come to our approach. As stated, $\xi_3$ and $\xi_4$ are fixed in terms of
$\gamma$, $\xi_1$, $\xi_2$, $a_s$ and $r_0$ by
eqs.(\ref{z3nnlo1s0}). As in the NLO case $\xi_1$ and $\xi_2$ are fixed by
eq.(\ref{z1nlo1s0}) and $\gamma=0$. As a result we have only one free
parameter, $\nu$. After performing the fit to the phase shifts one has the values:
\be
\begin{array}{llllll}
\gamma^\star=0\,\,\hbox{fm}^{-1}~,& \xi^\star_1=0.25~,& \xi_2^\star=0.03~, &
\xi_3^\star=0.21~, & \xi_4^\star=0.24~ &\nu=870 \hbox{ MeV}~,
\end{array}
\label{solonu}
\ee
corresponding to the solid line in the right panel of
fig.\ref{fig:1s0kswcm}, which is quite similar to the NNLO KSW one with two free
parameters although not so close to all the data points. In addition we show
by the dashed and short-dashed lines those curves that result from our
formalism keeping $\nu$ fixed at the values 500 MeV and 200 MeV,
respectively. The variation in the results from changes of $\nu$ is similar to
that obtained at the NLO, although the quality of the reproduction of data
have improved.

\begin{figure}[ht]
\psfrag{1S0(degrees)}{$S^1_0$; $\delta$(deg)}
\psfrag{1S0 (degrees)}{$S^1_0$; $\delta$(deg)}
\psfrag{p_cm(MeV)}{p (MeV)}
\psfrag{T_lab (MeV)}{T$_{lab}$ (MeV)}
\centerline{\epsfig{file=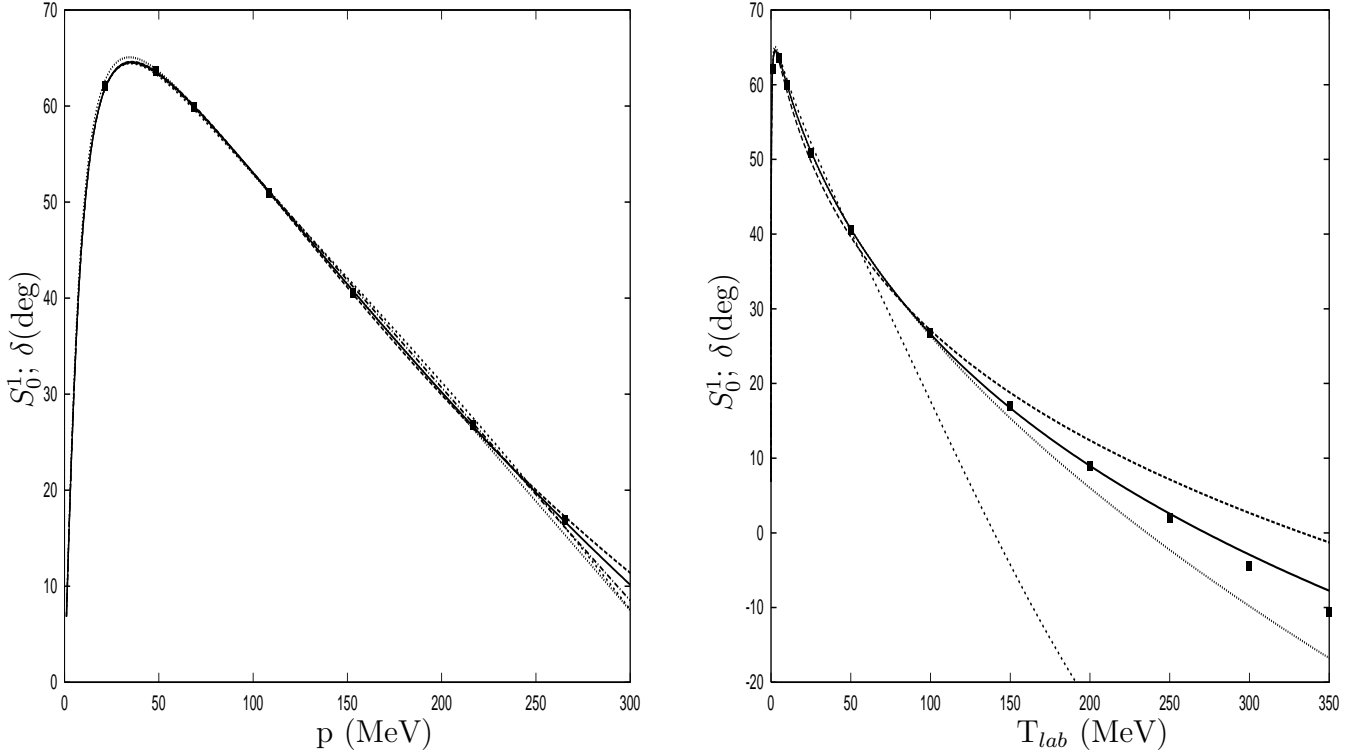,height=4.0in,width=7.0in,angle=0}}
\vspace{0.2cm}
\caption[pilf]{\protect \small
 Phase shifts for the $S^1_0$ channel. The dotted lines are the NNLO results
 from the pure KSW approach \cite{nnlo}. The rest of the curves are calculated
 within our scheme. Left panel: The different curves are hardly
 distinguishable. The short-dashed, dot-dashed, solid and dashed
 curves are our NNLO with $\nu$ taken as 200, 500, 700, and 900 MeV,
 respectively. $\xi_1$ and $\xi_2$ are taken as free parameters. The right
 panel corresponds to higher energies with the phase shifts as  function of
 $T_{lab}$. The solid line is the solid one of the left panel. The
 dashed line is the fit given in eq.(\ref{solonu}) at NNLO and the short-dashed one
 is the solid line of the left panel of fig.\ref{fig:1s0kswcm} calculated
 at NLO with $\nu$ fixed at 200 MeV. The data correspond to the Nijmegen
 partial wave analysis, ref.\cite{nij}.
\label{fig:1s0lab}}
\end{figure}

We now consider as well the possibility of fixing $\nu$ and then fitting
$\xi_1$ and $\xi_2$ while $\xi_3$ and $\xi_4$ are calculated once more from
eqs.(\ref{z3nnlo1s0}). We take the set of values $\nu=200$, 500, 700 and 900 MeV
that are shown in the left panel of fig.\ref{fig:1s0lab} by the short
dashed, dot-dashed, solid and dashed lines, respectively, although cannot be 
distinguished. In addition we also show the dotted line
corresponding to the NNLO results within the KSW approach from
ref.\cite{nnlo}. We see that basically there is no dependence on $\nu$
although the values of the fitted counterterms change substantially and they
are in order, for $\nu=200,$ 500, 700, 900 MeV: $\xi_1=-0.53$, 0.14, 0.56, 0.87 and 
$\xi_2=-0.21$, 0.04, 0.23, 0.36. On the other hand, the reproduction of data
is very good.

Let us now go to higher energies and show results up to laboratory kinetic
energies, $T_{lab}$=350 MeV (the threshold for pion production is at
$T_{lab}=280$ MeV). This is given in the right panel of
fig.\ref{fig:1s0lab}. The solid line corresponds to the solid one of the
left panel of the same figure, that is, when fixing $\nu=700$ MeV
and $\xi_1$ and $\xi_2$ taken as free parameters (as shown in the left panel of 
fig.\ref{fig:1s0lab} the change from one value of $\nu$ to another is almost negligible). In addition, the dashed line is the solid one of the right panel of 
fig.\ref{fig:1s0kswcm}, eq.(\ref{solonu}). The short 
dashed line is calculated at NLO within our approach with $\nu=200$ MeV 
(solid line in the left panel of fig.\ref{fig:1s0kswcm}) and the NLO KSW
results of ref.\cite{nnlo} are given by the dotted line.


\subsection{$S^3_1-D^3_1$ coupled channels}

We treat the counterterms  in analogous lines as described above for the 
$S^1_0$ channel, although now $\gamma$ and $\xi_5$ are taken as free
parameters. In addition, at NNLO in KSW there is a new parameter in this channel,
$\xi_6$, related to the $S^3_1-D^3_1$ mixing. One should take into account 
that although we use the same names for $\gamma$ and the $\xi$'s counterterms as in
the $S^1_0$ case, they are indeed different \cite{ksw,nnlo}. As before $\xi_1$ and 
$\xi_2$ are given at NLO in terms of $\gamma$, $a_s$ and $r_0$ from the 
ERE. At NNLO we then calculate $\xi_3$ and $\xi_4$ from $\xi_1$, 
$\xi_2$, $\gamma$, $a_s$ and $r_0$, independently of whether $\xi_1$ and $\xi_2$
are either taken 
as free parameters or fixed at NLO from ERE. The expressions for $\xi_1$, 
$\xi_2$, $\xi_3$ and $\xi_4$ determined from the ERE are the following:
\ba
\xi_1&=&\frac{1}{96 f^2m_\pi^2(-1+a_s\nu_1)^2\pi^2}\left\{
-g_A^2 M^2 (6\gamma^2-8 \gamma m_\pi+3m_\pi^2)(-1+a_s\nu_1)^2
+12 f^2 m_\pi^2(4-8 a_s \nu_1\right.\nn\\
&+&\left.a_s^2 M \nu_1^2\pi r_0)\right\}~,\nn\\
\xi_2&=&\frac{1}{32 f^2 m_\pi^2 \nu_1(-1+a_s \nu_1)\pi^2}\left\{
M(g_A^2 \gamma M(\gamma-2m_\pi)\nu_1 (-1+a_s\nu_1)+8f^2(-\nu_1^2\right.
\nn\\
&+&\left.
\gamma^2(-1+a_s\nu_1)+\gamma \nu_1(-1+a_s\nu_1))\pi)
\right\}~,
\label{erenlo3s1}
\ea
\ba
\xi_3&=&\frac{1}{2560 f^4 m_\pi^2 \nu_1^2(-1+a_s\nu_1)\pi^3}\left\{
80 f^2 g_A^2 M\nu_1(-1+a_s\nu_1)\pi(2\gamma^3 M+\gamma^2 M(-4m_\pi+\nu_1)
+8m_\pi^3 \nu_1 \pi \xi_2\right.\nn\\
&-&\left.2\gamma m_\pi \nu_1 (M+4m_\pi \pi \xi_2))-
640 f^4\pi^2 (\gamma^2 M\nu_1(1-a_s\nu_1)+\gamma^3(M-a_s M\nu_1)-
\gamma\nu_1(-1+a_s\nu_1)\right.\nn\\
&\times&\left.(M\nu_1-8m_\pi^2\pi \xi_2)+\nu_1^2
(M\nu_1+4m_\pi^2(-1+a_s\nu_1)\pi \xi_2))+g_A^4 M^3 \nu_1^2 (-1+a_s\nu_1)
(10\gamma^3-95\gamma^2 m_\pi\right.\nn\\
&-&\left.5\gamma m_\pi^2(-15+\log(4096))+
2m_\pi^3(-8+\log(67108864)))
\right\}~,\nn\\
\xi_4&=&-\frac{1}{107520 f^4 m_\pi^2 \nu_1(-1+a_s\nu_1)^2\pi^3}\left\{
13440 f^4 m_\pi^2\pi(8\gamma(-1+a_s\nu_1)^2(-1+2\pi^2\xi_1)+
\nu_1(-4+8\pi^2\xi_1\right.\nn\\
&+&\left.a_s^2\nu_1^2\pi(-Mr_0+8\pi\xi_1)+8a_s\nu_1
(1-2\pi^2\xi_1)))+1120 f^2 g_A^2 M(-1+a_s\nu_1)^2\pi(12\gamma^3 M+
\gamma^2M(-16 m_\pi\right.\nn\\
&+&\left.6\nu_1)+2\gamma m_\pi(M(3m_\pi-4\nu_1)+12m_\pi \nu_1 
\pi(\xi_1-2\xi_2))+m_\pi^2\nu_1(3M+8m_\pi \pi(-3\xi_1+4\xi_2)))\right.
\nn\\
&+&\left.
g_A^4M^3\nu_1(-1+a_s\nu_1)^2(1680\gamma^3-3325\gamma^2m_\pi-6m_\pi^3
(913+1248\log(2)+280\log(4)-336\log(16)\right.\nn\\
&-&\left.140\log(256))-840\gamma m_\pi^2
(-12+\log(256)))
\right\}~.
\label{erennlo3s1}
\ea

\begin{figure}[ht]
\psfrag{3S1 (degrees)}{$S^3_1$; $\delta$(deg)}
\psfrag{p_cm (MeV)}{p (MeV)}
\psfrag{Epsilon_1 (degrees)}{$\epsilon_1$ (deg)}
\centerline{\epsfig{file=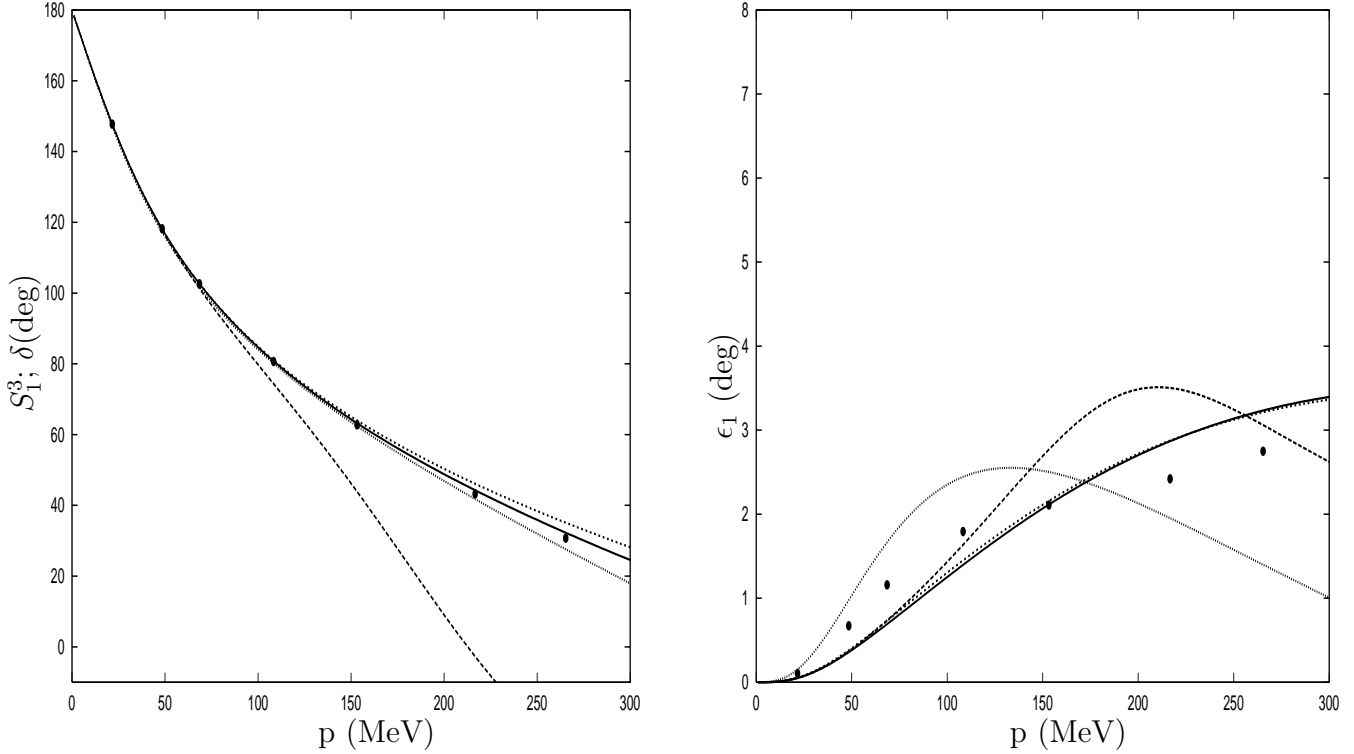,width=7.in,height=4.in}}
\vspace{0.2cm}
\caption[pilf]{\protect \small
 Phase shifts for the $S^3_1$ channel and mixing angle $\epsilon_1$. The
 dotted lines are the NLO KSW results \cite{nnlo}. The solid lines are the
 results from our approach corresponding to the fit of eq.(\ref{3s1nlobf}). In 
the  short-dashed lines the subtraction constants $\nu_1$
 and $\nu_2$ are fixed at 800 MeV. The dashed lines correspond to fixing $\nu_1$
 and $\nu_2$ at 200 MeV. The data come from the Nijmegen partial wave 
analysis, ref.\cite{nij}. 
\label{fig:3s1nlo}}
\end{figure}

At NLO the free parameters are $\gamma$, $\xi_1$, $\xi_2$, $\nu_1$ and
$\nu_2$ while in the pure perturbative treatment of ref.\cite{ksw,nnlo} the
latter two are absent. In ref.\cite{nnlo} $\gamma$ is fixed by requiring the
presence of the deuteron pole in the physical sheet and $\xi_2$ is given 
in terms of $\xi_1$ in order to avoid double poles.  As a result 
only one free
parameter remains at this order in the KSW scheme, $\xi_1$. This is
fitted to the low energy $S^3_1$ elastic phase shifts for $p\leq 80$ MeV in
ref.\cite{nnlo} with the resulting values \cite{nnlo}:
\be
\begin{array}{lll}
\gamma^\star=0.23 \,\,\hbox{fm}^{-1}~, & \xi_1=0.327~, & \xi_2^\star=-0.0936~,
\end{array} 
\label{ref:nlo3s1ksw}
\ee
where as usual the stars indicate that $\gamma$ and $\xi_2$ are not free
parameters in the fit. The results are given by the dotted line in fig.\ref{fig:3s1nlo}
both for the $S^3_1$ elastic phase shifts and for the $\epsilon_1$ mixing angle which is
defined such that $S_{11}=e^{2i\delta_{S^3_1}}\cos 2\epsilon_1$. The $D^3_1$
elastic phase shifts are presented in the fig.\ref{fig:d} and will be
discussed below together with the rest of D-waves. The agreement is 
remarkably good for the $S^3_1$ phase 
shifts and promising for the $\epsilon_1$ parameter, in the sense that the 
NNLO contributions are expected to improve the agreement with the
$\epsilon_1$ data. In the same figure we also present the resulting
curves from the scheme presented in this work at NLO. So as to reduce the number of
our free parameters as much as possible we impose that $\nu_2=\nu_1$, that is
quite natural if we think of the $\nu_i$ as coming from a cut-off as discussed
above. This constraint together with eqs.(\ref{erenlo3s1}),
which fix $\xi_1$ and $\xi_2$ in terms of the ERE parameters, $a_s=5.425$ fm
and $r_0=1.749$ fm, implies that only $\gamma$ and $\nu_1$ remain as free 
parameters.  We then obtain after the fit to the scattering data:
\be
\begin{array}{llll}
\gamma=0.41\,\,\hbox{fm}^{-1}~, & \xi_1^\star=0.33 ~, &\xi_2^\star=-0.06~,& 
\nu_2^\star=\nu_1=670 \hbox{ MeV}~.
\end{array}
\label{3s1nlobf}
\ee
The resulting curves are the solid ones in fig.\ref{fig:3s1nlo}. These curves
lie in general somewhat closer to data than those of the pure KSW treatment of
ref.\cite{nnlo} at NLO, particularly for cm three-momentum above 150
MeV. In addition we also consider the sensitivity of our results under a
change of the subtraction constants $\nu_2$ and $\nu_1$. The short-dashed
curves, the one lying highest in the left panel, correspond to fixing  
$\nu_2=\nu_1=800$ MeV and then $\gamma$ is fitted with the result 
$\gamma=0.41$ fm$^{-1}$ with $\xi_1^\star=0.32$ and $\xi_2^\star=-0.02$. 
Analogously, the lowest lying curve in the $S^3_1$ phase shifts, the dashed
ones, corresponds to taking $\nu_2=\nu_1=200$ MeV and then $\gamma=0.42$ fm$^{-1}$, 
$\xi_1^\star=0.44$ and
$\xi_2^\star=0.06$. Taking into account 
the lists of values for $\nu$ given in table \ref{tab:nu} as a function of an
hypothetical cut-off, it is clear that  a variation of the $\nu$'s from 200
MeV to 900 MeV, with the best fit at $\nu_2=\nu_1=670$ MeV, can be recast as
a large variation of $Q$, from around 300 MeV up to 1.6 GeV. It is also
remarkable the constancy of the value of $\gamma$ around $0.41$ fm$^{-1}$.

\begin{figure}[ht]
\psfrag{3S1 (degrees)}{$S^3_1$; $\delta$(deg)}
\psfrag{p_cm (MeV)}{p (MeV)}
\psfrag{Epsilon_1 (degrees)}{$\epsilon_1$ (deg)}
\centerline{\epsfig{file=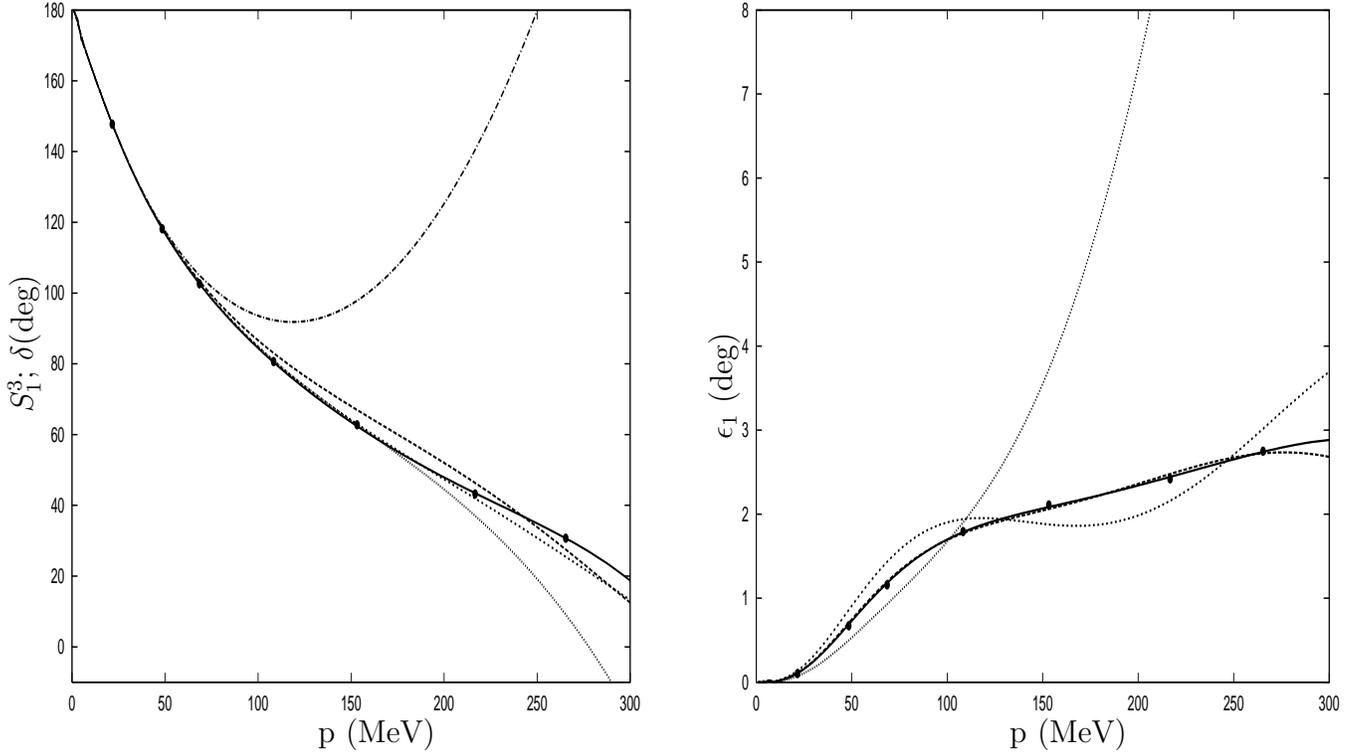,width=7.in,height=4.in}}
\vspace{0.2cm}
\caption[pilf]{\protect \small
 Phase shifts for the $S^3_1$ channel and mixing angle $\epsilon_1$. The
 dotted lines are the NNLO KSW results of ref.\cite{nnlo} with $\xi_5\neq 0$
 and the dashed-dotted line of the left panel when $\xi_5=0$. The
 solid lines correspond to the fit of eq.(\ref{3s1nnlobf}). The dashed ones
 are the fit of eq.(\ref{3s1nnlobfnu2nu1eq}). The short dashed lines
 correspond to the results of eq.(\ref{3s1nnlonu2fx}). For further details
 see the text.  The data are from the Nijmegen partial wave analysis, 
ref.\cite{nij}. 
\label{fig:3s1nnlo}}
\end{figure}

Now we come to the NNLO order results both at the perturbative level of
ref.\cite{nnlo} as well as from our scheme. As before, we consider first the pure
KSW analysis \cite{nnlo}. The value of $\gamma$ is the same as the
one at NLO eq.(\ref{ref:nlo3s1ksw}), since the pole in the perturbative
treatment comes only from $A_{-1}$, the leading contribution. Similarly as at NLO,
the number of free counterterms can be reduced by requiring the absence of
double and triple poles that appear in the NNLO KSW amplitudes. As a
result, $\xi_2$ and $\xi_3$ are expressed in terms of $\xi_1$ and
$\xi_4$. Thus the  free counterterms present in ref.\cite{nnlo} are finally
$\xi_1$, $\xi_4$, $\xi_5$ and $\xi_6$, which are fitted to the S-wave scattering
data, at this order the $D^3_1$ elastic phase shifts are free of any NNLO 
counterterm. The counterterm $\xi_6$ does not affect the elastic $S^3_1$ phase
shifts and only influences the $\epsilon_1$ mixing angle. The dotted lines in 
fig.\ref{fig:3s1nnlo} correspond to the pure KSW treatment with the 
set of values \cite{nnlo}:
\be
\begin{array}{llllll}
 \gamma^\star=0.23 \,\,\hbox{fm}^{-1}~, & \xi_1=0.432~, & \xi_2^\star=-0.0818~, & \xi_3^\star=0.165~, & \xi_4=0.399~, & \xi_5=0.26~ ,
\end{array} 
\label{ref:nnlo3s1ksw1}
\ee
by fitting the elastic $S^3_1$ phase shifts. The counterterms with a star
are not taken as free parameters in the fit as usual. The resulting phase shifts are
shown in the left panel of fig.\ref{fig:3s1nnlo} by the dotted line. In the 
second entry of ref.\cite{nnlo} only the mixing angle $\epsilon_1$ was 
considered and a fit 
was performed exclusively to the $\epsilon_1$ data from ref.\cite{nij}, without
considering the $S^3_1$ phase shifts. The new set of values are:
\be
\begin{array}{llll}
\gamma^\star=0.23 \,\,\hbox{fm}^{-1}~, & \xi_1=0.235~,& \xi_2^\star=-0.104~,
&\xi_6=0.385~.
\end{array}
\label{nnlo3s1ksw2}
\ee
The resulting $S^3_1$ phase shifts are much worse than those from the set of
eq.(\ref{ref:nnlo3s1ksw1}). Thus we only show the curves with the values of
eq.(\ref{ref:nnlo3s1ksw1}) and then we determine $\xi_6$ by performing a fit
to the $\epsilon_1$ data with the rest of counterterms fixed at the values of
eq.(\ref{ref:nnlo3s1ksw1}). We obtain then $\xi_6=0.50$ and the
resulting curve for $\epsilon_1$ is the dotted line shown in the right panel
of fig.\ref{fig:3s1nnlo}. As it is well known from the results of
ref.\cite{nnlo}, the NNLO results are worse than those at NLO already for 
$p\gtrsim 100$ MeV, despite the pion fields being  
explicitly included in the effective field theory. The
results shown in ref.\cite{nnlo} from the set of values of eq.(\ref{nnlo3s1ksw2})
give rise to the same divergent behavior as that indicated by the dotted lines of 
fig.\ref{fig:3s1nnlo}. It was also noted in ref.\cite{nnlo} that this bad
behavior  was due to large corrections from the twice iterated one pion
exchange diagrams which are enhanced by large numerical 
factors. In our power counting the input kernel, $\cR$, is infinitely 
iterated and with it the pion exchange  as any other contribution.

We now consider the results we obtain from our novel non-perturbative approach at
NNLO order in the expansion of ${\cal R}$. We follow the same treatment for
the counterterms as explained in the $S^1_0$ case, so that $\xi_1$, $\xi_2$ and
$\xi_3$, $\xi_4$ are fixed in terms of the ERE parameters at NLO
eq.(\ref{erenlo3s1}) and at NNLO eq.(\ref{erennlo3s1}),
respectively. Performing a fit to the scattering data, $S^3_1$ phase shifts
and $\epsilon_1$ mixing angle, up to $p\leq 300$ MeV, the values we obtain are:
\be
\begin{array}{lllllll}
\gamma=0.37\,\hbox{fm}^{-1}~, & \xi_1^\star=0.44~, & \xi_2^\star=0.01~, & 
\xi_3^\star=0.05~, & \xi_4^\star=0.04~, & \xi_5=0.19~, & \xi_6=0.58~,\\
 \nu_1=190\hbox{ MeV}~, & \nu_2=620 \hbox{ MeV}~. & & & & &
\end{array}
\label{3s1nnlobf}
\ee
The results of this fit are presented in the left and right panels of
fig.\ref{fig:3s1nnlo} by the solid lines. As we see the agreement with data is
remarkably good. We now show the sensitivity of the results under changes of the
subtraction constants $\nu_1$ and $\nu_2$. For that, we first impose the
constraint $\nu_2=\nu_1$ as in the NLO case, and perform the fit. The fitted
parameters are:
\be
\label{3s1nnlobfnu2nu1eq}
\begin{array}{lllllll}
\gamma=0.41\,\hbox{fm}^{-1}~, & \xi_1^\star=0.34~, & \xi_2^\star=0.00~, & 
\xi_3^\star=0.06~, & \xi_4^\star=0.16~, & \xi_5=0.25~, & \xi_6=0.51~,\\
 \nu_1=500\hbox{ MeV}~, & \nu_2^\star=500 \hbox{ MeV}~. & & & & &
\end{array}
\ee
the results are shown in fig.\ref{fig:3s1nnlo} by the dashed lines. They are
quite similar to those of the fit of eq.(\ref{3s1nnlobf}) and essentially
identical for the $\epsilon_1$ mixing angle. Let us note that $\nu_1$ has
changed very appreciably with respect to eq.(\ref{3s1nnlobf}) and the fit
continues being rather acceptable.
Finally, in order to have as well a large variation of $\nu_2$ from the value
given in eq.(\ref{3s1nnlobf}), we fix $\nu_1=600$ MeV and $\nu_2=200$ MeV and fit $\gamma$,
$\xi_5$, $\xi_6$. The resulting values are:
\be
\begin{array}{lllllll}
\gamma=0.55\,\hbox{fm}^{-1}~, & \xi_1^\star=0.33~, & \xi_2^\star=0.10~, & 
\xi_3^\star=-0.19~, & \xi_4^\star=-0.11~, & \xi_5=0.20~, & \xi_6=0.57~,\\
 \nu_1^\star=600\hbox{ MeV}~, & \nu_2^\star=200 \hbox{ MeV}~. & & & & &
\end{array}
\label{3s1nnlonu2fx}
\ee
The results correspond to the short dashed lines. The $S^3_1$ phase shifts are
very similar to those of the best fit of eq.(\ref{3s1nnlobf}) although the
$\epsilon_1$ mixing angle curve is more different than that obtained from the
fit with $\nu_2=\nu_1$, eq.(\ref{3s1nnlobfnu2nu1eq}). Thus, we see that 
the dependence on the subtraction constants $\nu_1$ and $\nu_2$ is rather mild 
and can be reabsorbed to a large extend in new values of the KSW counterterms, 
$\gamma$ and  $\xi_i$.

\begin{figure}[ht]
\psfrag{3S1 (degrees)}{$S^3_1$; $\delta$(deg)}
\psfrag{T_lab (MeV)}{T$_{lab}$ (MeV)}
\psfrag{Epsilon_1 (degrees)}{$\epsilon_1$ (deg)}
\centerline{\epsfig{file=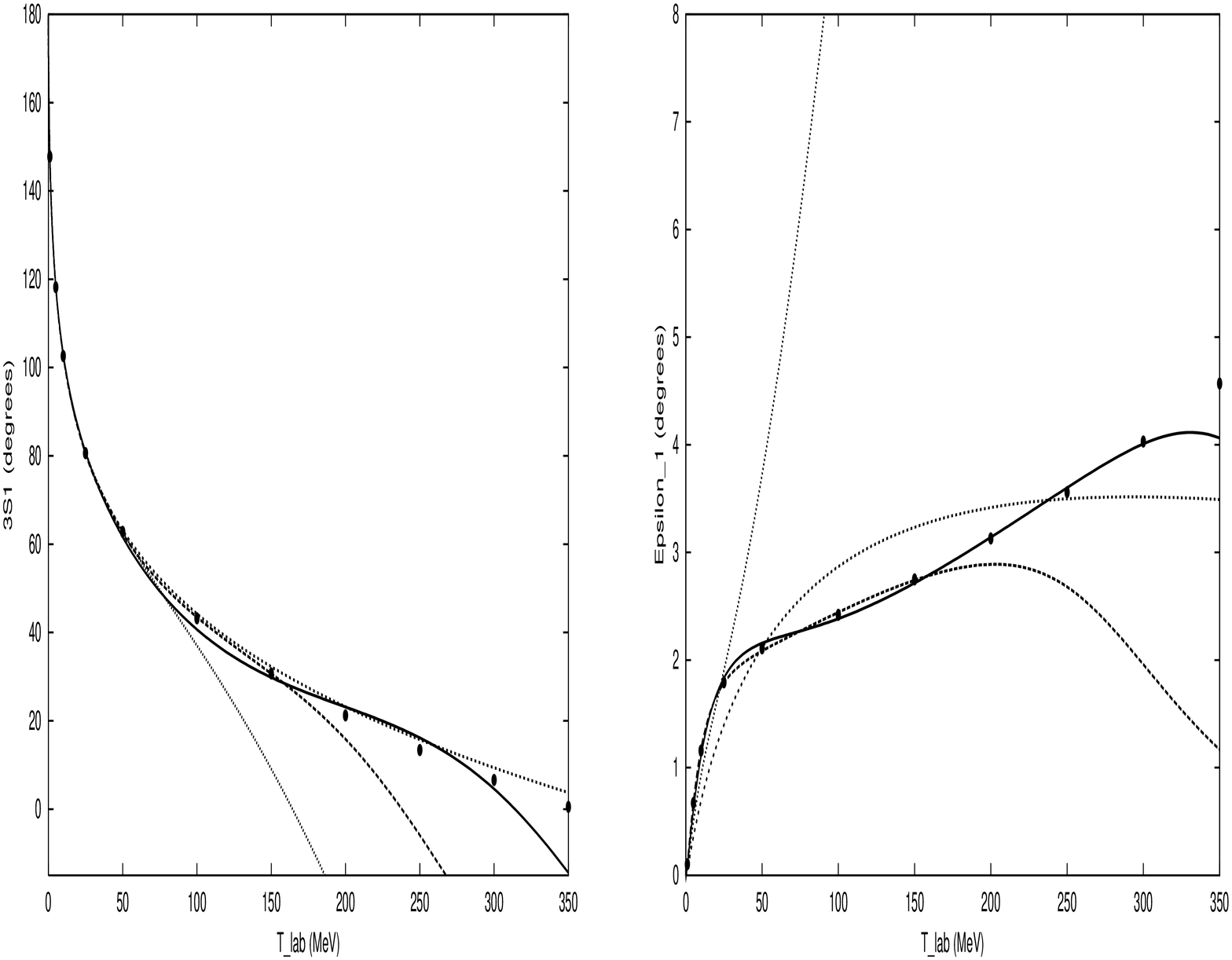,width=7.in,height=4.in}}
\vspace{0.2cm}
\caption[pilf]{\protect \small
 Phase shifts for the $S^3_1$ channel and mixing angle $\epsilon_1$. The
 dotted lines are the NNLO KSW result  \cite{nnlo} with $\xi_5\neq 0$. The
 short dashed lines represent the NLO results from eq.(\ref{3s1nlobf}). The
 dashed lines are the NNLO fit to data for $p\leq 300$ MeV given in 
eq.(\ref{3s1nnlobf}). The solid lines are the new fit of eq.(\ref{3s1nnlobf400})
 for $p\leq 400$ MeV. The data
  correspond to the Nijmegen partial wave analysis, ref.\cite{nij}. 
\label{fig:3s1}}
\end{figure} 

In the KSW approach, when solving the renormalization group 
equations one has \cite{nnlo},
\be
\label{xi5}
\xi_5=\rho\frac{m_\pi^2 M}{4\pi}~
\ee
and since $\rho$ is order $p^0$, given by $\Lambda_{NN}$, then $\xi_5$ is 
formally a quantity of order $p^2$. Hence, within the KSW scheme its 
contributions to the scattering amplitude are order $p^2$ and start at
N$^3$LO. This situation is the standard one in ref.\cite{nnlo} and the results 
from the KSW perturbative treatment with $\xi_5=0$, with the rest of
counterterms $\xi_1-\xi_4$ given in eq.(\ref{nnlo3s1ksw2}), correspond to the
dashed-dotted line in fig.\ref{fig:3s1nnlo}, the one that lies above all the
other ones in the left panel. We see then that the elastic 
$S^3_1$ phase shifts diverge for cm three-momenta much smaller than for 
the $\xi_5\neq 0$ case but similar to those cm three-momenta  where the
divergence starts for the $\epsilon_1$ mixing angle, which indeed 
is independent of $\xi_5$ at NNLO in the KSW approach. 

Now, let us discuss why 
in our approach we should take $\xi_5$ as a free parameter without taking into
 account eq.(\ref{xi5}). The point is the following. The order of the
counterterms in KSW, in particular those given rise to $\xi_5$, are determined
by comparing with the effective range expansion in the pionless effective field
theory  once the power divergence subtraction (PDS) scheme is adopted
\cite{ksw}. This comparison is exact, even if the series are truncated, since
both of them implies an expansion in powers of the cm three-momentum and the
comparison is performed order by order. In this process one formally books the
scattering length as order $p^{-1}$ and the shape parameters $r_n$ as order
zero, proportional to $1/\Lambda_{NN}$. Once this is performed, those 
operators that enter in the Lagrangian up to some order can be determined, and
then the corresponding KSW partial wave amplitudes up to the same order can be
calculated as well. As we have
already explained, these amplitudes are taken as the input of our approach in
order to fix ${\cal R}$ by the matching process outlined above in 
sec.\ref{sec:form}. In the pure KSW treatment one requires  the amplitudes
being independent under changes of the subtraction point so that the
renormalization group equations (RGE) follow for the different local 
counterterms. In
practice, even when pions are removed and only local operators remain, this
 guarantees that the scattering partial wave is subtraction point
independent only perturbatively, in the sense that this only happens when the 
Schr\"odinger equation is solved perturbatively.\footnote{This acquires a clear
  meaning if one thinks of the pionless effective field theory, where to solve
  the Schr\"odinger equation is a trivial task.} But this is 
precisely the point we want to avoid in our
formalism so as to take care properly of the large $2M/p$ factors that
appear from the two nucleon intermediate states in the unitarity bubbles. At this
point our formalism is an hybrid between that of Weinberg \cite{law} and the 
KSW formalism \cite{ksw}. On the one hand, we perform an expansion of an
interacting kernel ${\cal R}$ as in ref.\cite{law} but on the other we do
that by considering directly the scattering amplitudes as in ref.\cite{ksw}. 
Indeed, there is a residual dependence in our amplitudes on the subtraction 
constant $\nu$, expected to correspond to higher order operators in the KSW 
Lagrangian, similarly as in the Weinberg approach where there is cut-off 
dependence that is expected to become softer as the order of the calculation 
increases, as also occurs in our case as discussed above. 
To sum up, our scheme makes use of the KSW power counting  in order to obtain
the Lagrangian, employs
Feynman diagrams to calculate from this Lagrangian the KSW amplitudes at some
order (where the order of any combination of counterterms is determined
directly from the ones of the original counterterms in the Lagrangian, without
making use of perturbative RGE arguments). Then, these amplitudes are matched 
with the general expression of eq.(\ref{keyt}) so that the  interacting kernel 
${\cal R}$ is determined. Thus, as a result, one ends with partial wave amplitudes with 
the unitarity bubbles resummed as in Weinberg's scheme but in an
analytical way based on unitarity and analyticity.

\begin{figure}[ht]
\psfrag{3S1 (degrees)}{$S^3_1$; $\delta$(deg)}
\psfrag{p_cm (MeV)}{p (MeV)}
\psfrag{Epsilon_1 (degrees)}{$\epsilon_1$ (deg)}
\centerline{\epsfig{file=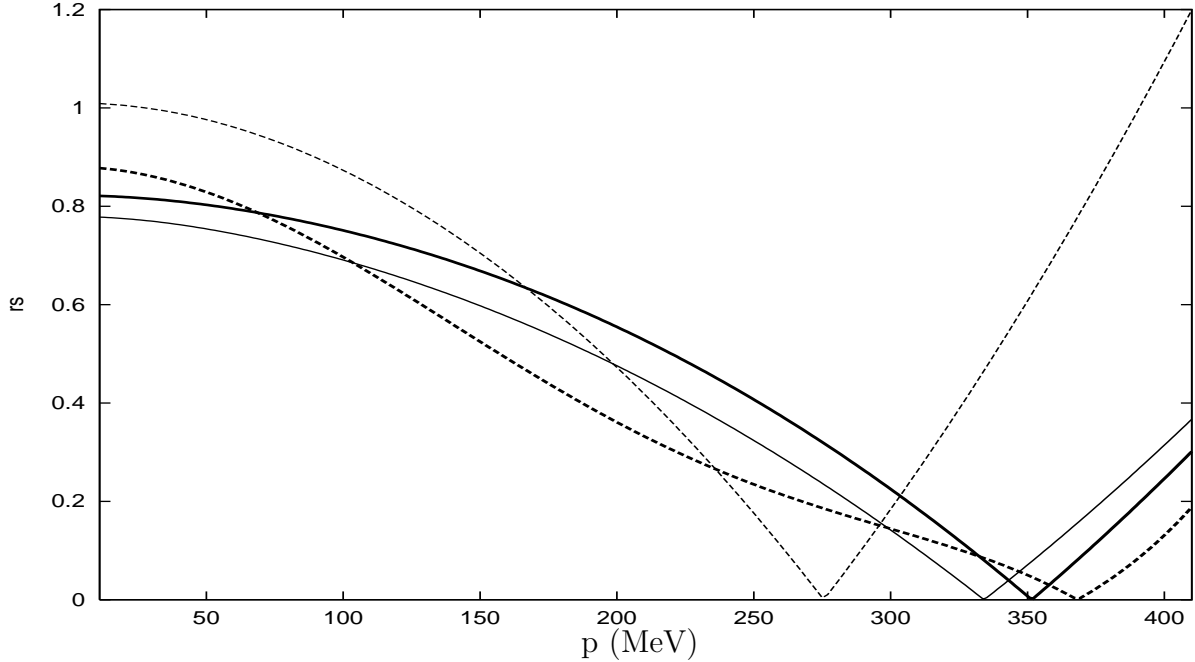,width=3.5in,height=6.5in,angle=-90}}
\vspace{0.2cm}
\caption[pilf]{\protect \small
 $|{\cal R}^{NLO}|$ (thin solid line), $|{\cal R}^{NNLO}|$ (thick solid line)
 for the $S^1_0$ channel and $|R_{11}^{NLO}/4|$ (thin dashed line) and
 $|R_{11}^{NNLO}/4|$ (thick dashed line) for the $S^3_1$ channel, as a function
  of the cm three-momentum.
\label{fig:rs}}
\end{figure}

Encouraged by the rather good fits obtained from our approach up to $p\lesssim
300$ MeV as depicted in figs.\ref{fig:3s1nlo} and \ref{fig:3s1nnlo}, we now show
in fig.\ref{fig:3s1} our results for higher energies, namely for  $T_{lab}\lesssim 350$ MeV, as 
similarly shown  in the studies of nucleon-nucleon scattering within the Weinberg
approach in ref.\cite{epe1}. Let us note that for $p=300$ MeV one has
$T_{lab}\simeq 190$ MeV. We present in fig.\ref{fig:3s1} several lines. The
dashed lines correspond to the solid ones of fig.\ref{fig:3s1nnlo}, with the
parameters given in eq.(\ref{3s1nnlobf}). We see, as
already shown in fig.\ref{fig:3s1nnlo}, that the agreement with data for
$T_{lab}\lesssim 190$ MeV is very good but from $T_{lab}=200$ MeV starts deviating 
from data.  The dotted lines are  
the NNLO KSW results from ref.\cite{nnlo} with $\xi_5\neq 0$, already presented in
fig.\ref{fig:3s1nnlo} with the same type of line. We also show with the
short-dashed lines our NLO results given in eq.(\ref{3s1nlobf}) with only two
free parameters in the fit. We see that they follow closely the trend of the
$S^3_1$ phase shifts in all the energy interval although the agreement is not
so good for the $\epsilon_1$, similarly as also happens in the KSW treatment
at NLO. Finally by the solid line we present a new fit up to $p\leq 400$ MeV,
where $\xi_1$ and $\xi_2$ are fitted, instead of begin fixed by the ERE at
NLO, eqs.(\ref{erenlo3s1}). On the other hand, $\xi_3$ and $\xi_4$ continue to
be given in terms of $\xi_1$, $\xi_2$ and $\gamma$ from
eqs.(\ref{erennlo3s1}). The resulting parameters are:
\be
\begin{array}{lllllll}
\gamma=0.51\,\,\hbox{fm}^{-1}~, & \xi_1=0.27~, & \xi_2=0.06~, &
\xi_3^\star=-0.10~, & \xi_4^\star=-0.05~, & \xi_5=0.17~,& \xi_6=0.46~,\\
\nu_1=180 \hbox{ MeV}~,& \nu_2=560 \hbox{ MeV}~. & & & & & 
\end{array}
\label{3s1nnlobf400}
\ee
The reproduction of the data is quite good, specially for the mixing angle
$\epsilon_1$. Nevertheless, we observe some deviation from data in the low
energy region around $T_{lab}\simeq 100$ MeV ($p\simeq 215$ MeV) which does not
appear in the other fits obtained by fitting data only up to $p=300$ MeV or
$T_{lab}\lesssim 190$ MeV, that is for lower energies. This can be interpreted 
as an indication that we are forcing too much the capability of our 
approach accordingly with an estimated
$\Lambda_{NN}$ around 400 MeV. This is also supported by the divergence of
the fit of eq.(\ref{3s1nnlobf}) for $T_{lab}$ above 200 MeV, solid lines in 
fig.\ref{fig:3s1nnlo} and dashed ones in fig.\ref{fig:3s1}, as well as from the
results shown in fig.\ref{fig:rs} which are discussed in the next
paragraph. In this sense the
fit given in eq.(\ref{3s1nnlobf400}) is just a way to fit data up to
rather high energies. Nonetheless, it is reassuring that the values of the
parameters, although there are now two free parameters more, are very 
similar to those obtained above for lower energies, 
e.g. in eq.(\ref{3s1nnlobf}).

We also show in fig.\ref{fig:rs} the absolute values $|\cR^{NLO}|$ and 
$|\cR^{NNLO}|$ for the $S^1_0$ channel (solid lines) and the same, although 
divided by four to keep the lines on the same scale, for the matrix element
$|R_{11}|$  of the $S^3_1$ channel. The thick lines
refer to the NNLO calculations and the thin ones to NLO. 
The convergence properties are quite good in a broad range of the center of mass 
three momentum $p$, with strong divergences for $p$ around 
$\Lambda_{NN}\simeq 400$ MeV. This scale is 
the one expected for the KSW EFT \cite{ksw}, although in the end this EFT does not 
converge in the triplet channels for $p\gtrsim 100$ MeV, as already discussed. 
We see that within 
our scheme, at the same time that we keep the KSW power counting, we are able to fulfill 
these expectations. This is the main aim of the present investigation.

\subsection{$P$, $D$, $F^3_2$ and $G^3_3$ partial waves}

\begin{figure}[ht]
\psfrag{1P1}{$P^1_1$; $\delta$(deg)}
\psfrag{3P0}{$P^3_0$; $\delta$(deg)}
\psfrag{3P1}{$P^3_1$; $\delta$(deg)}
\psfrag{3P2}{$P^3_2$; $\delta$(deg)}
\psfrag{3F2}{$F^3_2$; $\delta$(deg)}
\psfrag{ep2}{$\epsilon_2$; $\delta$(deg)}
\psfrag{3D1 (degrees)}{$D^3_1$; $\delta$(deg)}
\psfrag{T_lab (MeV)}{$\begin{array}{c}\\T_{lab} \hbox{(MeV)}\end{array}$}
\centerline{\epsfig{file=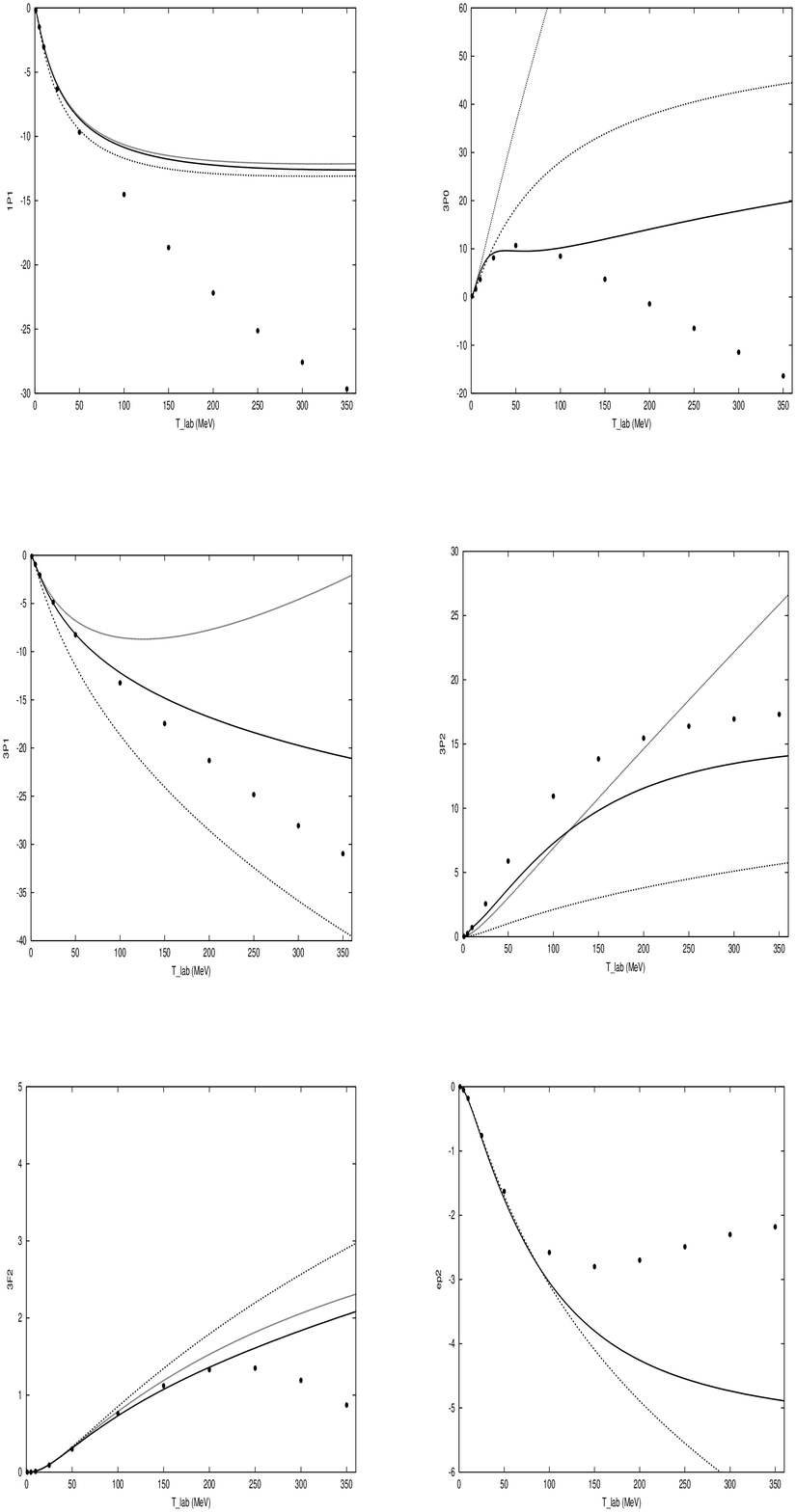,width=7.5in,height=7.5in}}
\vspace{0.2cm}
\caption[pilf]{\protect \small
 Phase shifts for the  $P^1_1$, $P^3_0$, $P^3_1$, $P^3_2$, $F^3_2$ and 
 $\epsilon_2$ from left to right and top to bottom, respectively. The dotted line, 
 when present, is the NNLO KSW result 
 \cite{nnlo}. The dashed line represents the NLO results from eq.(\ref{keyt}). 
 The solid lines are the  NNLO results. The data correspond to the Nijmegen 
 partial wave analysis, ref.\cite{nij}.
\label{fig:p}}
\end{figure}

\begin{figure}[ht]
\psfrag{1D2}{$D^1_2$; $\delta$(deg)}
\psfrag{3D1}{$D^3_1$; $\delta$(deg)}
\psfrag{3D2}{$D^3_2$; $\delta$(deg)}
\psfrag{3D3}{$D^3_3$; $\delta$(deg)}
\psfrag{3G3}{$G^3_3$; $\delta$(deg)}
\psfrag{ep3}{$\epsilon_3$; $\delta$(deg)}
\psfrag{3D1 (degrees)}{$D^3_1$; $\delta$(deg)}
\psfrag{T_lab (MeV)}{$\begin{array}{c}\\T_{lab} \hbox{(MeV)}\end{array}$}
\centerline{\epsfig{file=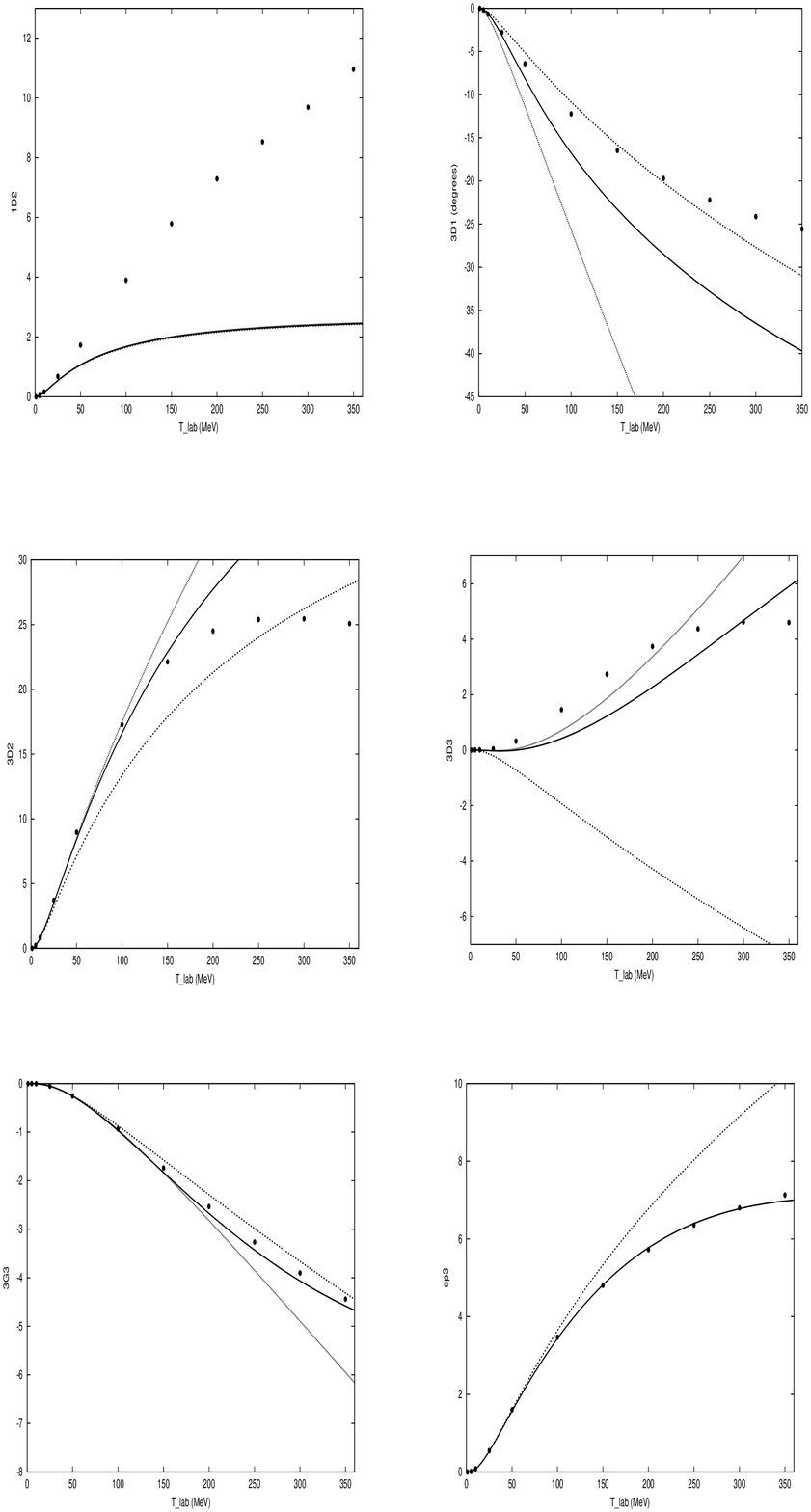,width=7.5in,height=7.5in}}
\vspace{0.2cm}
\caption[pilf]{\protect \small
Phase shifts for the $D^1_2$, $D^3_1$, $D^3_2$, $D^3_3$, $G^3_3$ and 
 $\epsilon_3$  from left to right and top to bottom, respectively. The dotted line, 
 when present, is the NNLO KSW result 
 \cite{nnlo}. The dashed line represents the NLO results from eq.(\ref{keyt}). 
 The solid lines are the  NNLO results. The data correspond to the Nijmegen 
partial wave analysis, ref.\cite{nij}.
\label{fig:d}}
\end{figure}

Up to NNLO in the KSW approach, order $p$, the physics included in the 
description of partial waves higher than $S$ waves, except for the 
mixing between the $S^3_1-D^3_1$ ones,  consists only of the one pion exchange, 
order $p^0$, and the reducible part of the twice iterated one pion exchange,
order $p$. While the former contribution constitutes the LO one of the Weinberg's
scheme \cite{kolck,epe1}, the latter is generated by solving the
Lippmann-Schwinger equation. Hence,  the
NNLO results in the KSW approach are included in the leading ones of Weinberg's
scheme. It is clear then that we are neglecting important contributions for
the higher partial waves, particularly for the $P$ and 
$D$ waves, since from the results of the Weinberg's approach we know that two
pion exchange irreducible diagrams, counterterms for the $P$-waves 
at order $p^2$, as well as the $\pi N$ counterterms (which are
 saturated to a large extend by the $\Delta$ isobar) are important
 contributions that would rise in higher order calculation within the KSW,
 beyond NNLO. Hence a full N$^3$LO calculation from the KSW Lagrangian should be 
pursued and then used within our approach to 
offer a more complete study of these higher partial waves. For the $F$ and
 $G$ waves and mixing parameters $\epsilon_2$ and $\epsilon_3$, pion exchange 
dominates and the aforementioned extra 
contributions are not so important, see e.g.\cite{kaiser}. This is 
also clear from our results given in figs.\ref{fig:p} and \ref{fig:d}. At the order 
we are working in these partial waves there are no counterterms and the only 
free parameters are the subtraction constants $\nu$'s, one for each partial wave.  
In most of the channels, they take  arbitrarily large or negative 
values, that is, any value with modulus typically above 
$\Lambda_{\chi pT}\simeq 0.7-1$ GeV gives essentially the same results. 
Our curves are indeed quite similar 
to those obtained in the Weinberg's approach at LO, 
see \cite{epe1}. Our results at NNLO, solid lines, improve those of 
the NLO, dashed lines, except for the $D^3_1$ where, though the NNLO is better at 
low $T_{lab}$, they depart from data more than the NLO ones for 
higher energies. They also improve the results from the pure KSW amplitudes, dotted 
lines, although in these cases, as expected, the resummation effects are not 
so important as in the S-waves. 

Since most of the subtraction constants $\nu$ for the analyzed $P$, $D$, $F$ and $G$ 
waves take arbitrarily large absolute values we would like to show that in the limit when
$|\nu|\rightarrow \infty$ our approach  reduces to the
Inverse Amplitude Method \cite{iam}. This method would 
consist of expanding the inverse of a partial wave, following in our present
problem the KSW power counting, 
 and then calculate the partial wave by inverting exactly the
previous expansion. For instance,  we have given in eq.(\ref{kswexp}) the expansion 
of the inverse of the elastic $S^1_0$ partial wave  within the
KSW power counting (in ref.\cite{iam} one uses for meson-meson scattering the standard
chiral perturbation theory counting and for pion-nucleon processes the chiral 
perturbation theory counting is supplied with the heavy baryon one). Then the inverse 
amplitude method would imply:
\be
T_{S^1_0,S^1_0}=\left[\left(\frac{1}{A_{-1}}\right)-
\left(\frac{A_0}{A_{-1}^2}\right)
+\left(\frac{A_0^2-A_1 A_{-1}}{A_{-1}^3}\right)\right]^{-1}~.
\label{eqiam}
\ee
Its generalization to coupled channels is straightforward by making use of a
matrix notation. For example in the $S^3_1-D^3_1$ coupled channels one should
just invert exactly the matrix $(A^{KSW}_{S^3_1-D^3_1})^{-1}$ whose matrix
elements are given in eqs.(\ref{ksw3sdexp}), and so on for any other partial waves.
This kind of results are also usually referred as Pad\'e resummations.

Let us see that  our formalism reduces to
the inverse amplitude method  when $|\nu|\rightarrow \infty$. Consider first
the elastic case. Then, the contribution of order $p^i$ in the KSW power
counting, from the expansion of $1/{\cal R}+g$ in eq.(\ref{keyt}), has one
piece involving the $R_i$ component of ${\cal R}$. Isolating $R_i$ one has:
\be
\label{ri}
R_i=-R_0^2\left[\left(1/A\right)_i+g_i\right]~+...,
\ee
where $g_i$ is now the contribution of the $g$ function of order $p^i$ and $(1/A)_i$ 
is that of the 
inverse of the KSW partial wave. Since $R_0$
scales as $1/\nu$, see eq.(\ref{exp}) for the case of large scattering lengths
and (\ref{expnatu}) for the case of natural ones, then $R_i$ has a
contribution scaling as $1/\nu^2$. This is the only contribution for $i=1$ and
then $R_1$ scales as $1/\nu^2$. Let us demonstrate by induction that each 
$R_i$ with $i\geq 1$ scales as $1/\nu^2$, while $R_0$
scales only as $1/\nu$. To show that, let us consider the dots which indicate 
other terms from the
expansion of $1/{\cal R}$ involving $R_m$ with $1\leq m\leq i-1$. We can write
these contributions as:
\be
\sum_{m=1}^{i-1}C_{k_1\,k_2\,...\,k_m}^{a_1\,a_2\,...\,a_m}\frac{R_{k_1}^{a_1}
  R_{k_2}^{a_2}... R_{k_m}^{a_m}}{R_0^{a_1+a_2+...+a_m-1}}~,
\label{sumri}
\ee
where $k_p$ ($1\leq p\leq i-1$) indicates the order of the corresponding
factor $R_{k_p}$, in addition all the $k_p$ are different from each other and
fulfill $a_1 k_1+a_2 k_2+...+a_m k_m=i$. Because each of the $R_{k_p}$ in
eq.(\ref{sumri}) scales at least
as $1/\nu^2$ one can count easily the dominant power of $\nu$ from
eq.(\ref{sumri}) when $|\nu|\rightarrow \infty$. One simply has:
\be
\frac{\nu^{a_1+a_2+...+a_m-1}}{\nu^{2(a_1+a_2+...+a_m)}}=\frac{1}{\nu^{a_1+a_2+...+a_n+1}}~.
\ee
But the sum $a_1+a_2+...+a_n+1\geq 3$. To show this, let us consider first
those terms in the sum of eq.(\ref{sumri}) with all the factors the same 
and equal to $R_{k_1}$. It follows
then that $a_1\geq 2$ since $1\leq k_1\leq i-1$. The rest of terms will have
at least two different factors, then
there must be two $a'$s, let us say $a_1$ and $a_2$, different from zero and every
one $\geq1$. Then the scaling of $R_i$ in $\nu$ when $|\nu|\rightarrow \infty$
is dominated the one from eq.(\ref{ri}) and goes like $1/\nu^2$.

Let us consider the expansion of  $1/{\cal R}+g$ in eq.(\ref{keyt}) in powers
of $1/\nu$ in the limit $|\nu|\rightarrow \infty$ with ${\cal R}$ determined
up to order $p^n$ in the KSW counting. We also write for $g$ only the sum
$\sum_{k=0}^n g_k$ since higher order terms are just relativistic corrections
and numerically are negligible:
\be2
\label{expkeyt}
T_{L^{2S+1}_J,L^{2S+1}_J}=-\left(\frac{1}{R_0+\sum_{m=1}^n R_m}+\sum_{k=0}^n
  g_k \right)^{-1}=
-\left(\frac{1}{R_0}-\sum_{m=1}^n \frac{R_m}{R_0^2}+{\cal O}(\nu^{-1})
+\sum_{k=0}^n g_k \right)^{-1}~.
\ee
Where we neglect the ${\cal O}(\nu^{-1})$ terms compared to those explicitly shown
in the equation above which are ${\cal O}(\nu^0)$ and are the only ones that survive 
in the limit $|\nu|\rightarrow \infty$. Thus, as a result of the
matching process detailed in sec.\ref{sec:form} to obtain the $R_i$ functions
that we apply now for $|\nu|\rightarrow \infty$, one has between the brackets of 
eq.(\ref{expkeyt}) the 
expansion of the inverse of the KSW amplitude up to order $p^n$. So that the
resulting $T_{L^{2S+1}_J,L^{2S+1}_J}$ partial wave from eq.(\ref{keyt}) for 
$|\nu|\rightarrow \infty$ is the one given by the
inverse amplitude method eq.(\ref{eqiam}). For finite $\nu$'s this is no
longer the case.  The neglected terms  of order $\nu^{-1}$ and
higher in eq.(\ref{expkeyt}) guarantee the matching with the expansion of the
inverse the KSW amplitudes up to order $p^n$ and also give rise to higher
order terms beyond those considered in the matching process.  It is only
necessary a little thought to extend this discussion for the elastic case to the 
coupled channel one by making use of a matrix language.
 
\section{Conclusions}
We have established a new analytic expansion in order to treat systematically 
nucleon-nucleon interactions. Analyticity and unitarity properties of the inverse of a
partial wave amplitude are used to resum the unitarity bubbles in terms of an
on-shell interacting kernel ${\cal R}$ and the unitarity loop function $g$. 
 The former is determined by performing an
expansion with the KSW power counting of the aforementioned general expression
for a partial wave and by 
matching this expansion with that of the inverses of the pure KSW amplitudes
up to a definite order. In practice we have performed this matching with
NLO and NNLO KSW amplitudes \cite{ksw,nnlo}. As a result, a hybrid scheme emerges
that treats pions non-perturbatively but in harmony with the KSW power counting, and  
that restores the expected range of convergence of the KSW EFT about
$\Lambda_{NN}=400$ MeV. It reminds also the Weinberg's approach \cite{law}
in that a perturbative expansion is performed for an interacting kernel
instead of making it directly to the partial waves. In our case
this interacting kernel is ${\cal R}$ while in Weinberg's approach is the
potential, $V$. However, it is worth stressing that while $\cR$ is on-shell, $V$ is
off-shell and this is an important point that allows our formalism to be
purely analytic and fairly more simple than that of ref.\cite{law}.
It is also important to remark that for the S-waves we have achieved a very good agreement with
data, both for the phase shifts as well as for the mixing angle $\epsilon_1$,
up to the opening of the $NN\pi$ threshold. 

Further applications of the present scheme should be pursued, particularly to
three body problem where the disposal of analytical methods in the two-nucleon
sector is well worth \cite{three} and furthermore a  KSW N$^3$LO calculation 
should also be performed within our scheme particularly for a more complete study of 
the $P$ and $D$ partial waves. Finally, another issue is to extend this
knowledge of the nucleon-nucleon interactions in vacuum to nuclear matter \cite{oller1,oller2}.
\label{sec:con}
\def\theequation{\arabic{section}.\arabic{equation}}
\setcounter{equation}{0}

\vspace{1cm}
\noindent {\bf Acknowledgments}

\medskip
I would like to thank E. Epelbaum for interesting discussions that we 
held in J\"ulich on nucleon-nucleon interactions. I am 
also grateful to the Benasque Center for Science, where part of this work
was done, for its support and enjoyable atmosphere. This work is partially
supported by the DGICYT project FPA2002-03265.

\end{document}